\documentclass[useAMS,usenatbib]{mn2e}
\usepackage{float}
\usepackage{rotating}
\usepackage{color}
\usepackage{amsmath}
\usepackage{amssymb}
\usepackage{multirow}
\usepackage{multicol}
\usepackage{url}
\usepackage{natbib}
\usepackage{bm}

\newcommand{\kms}{\ensuremath{{\rm km\,s}^{-1}}}
\newcommand{\msun}{\ensuremath{{\rm M}_{\odot}}}

\newcommand{\note}[1]{}

\newcommand{\tw}{\columnwidth}
\def\mean#1{\left< #1 \right>}

\newcommand{\Nc}{\ensuremath{{\cal N}_{\rm o}}}
\title[Two paths of cluster evolution]{Two paths of cluster evolution: global expansion versus core collapse}

\author[O'Leary, Stahler, \& Ma]{Ryan M.\ O'Leary\thanks{E-mail: oleary@berkeley.edu} ,  Steven W.\  Stahler, \& Chung-Pei Ma\\
Department of Astronomy and Theoretical Astrophysics Center, University of California, Berkeley, CA 94720, USA\\
}
\begin{document}
\maketitle

\begin{abstract}

All gravitationally bound clusters expand, due to both gas loss from
their most massive members and binary heating. All are eventually
disrupted tidally, either by passing molecular clouds or the
gravitational potential of their host galaxies.  However, their
interior evolution can follow two very different paths.
 Only clusters of sufficiently large initial population and size
undergo the combined interior contraction and exterior expansion that
leads eventually to core collapse. In all other systems, core collapse
is frustrated by binary heating.
 These clusters globally expand
for their entire lives, up to the point of tidal disruption.

Using a suite of direct $N$-body calculations, we trace the ``collapse line'' in $r_v-N$
space that separates these two paths. Here, $r_v$ and $N$ are the
cluster's initial virial radius and population, respectively. For
realistic starting radii, the dividing $N$-value is from $10^4$ to
over $10^5$.
We also show
that there exists a minimum population, $N_{\rm min}$, for core collapse.
Clusters with $N < N_{\rm min}$ tidally disrupt before core collapse occurs. At the Sun's
Galactocentric radius, $R_G = 8.5$ kpc, we find $N_{\rm min} \gtrsim 300 $. The minimum
population scales with Galactocentric radius as $R_G^{-9/8}$.

The position of an observed cluster relative to the collapse line can
be used to predict its future evolution. Using a small
sample of open clusters, we find that
most lie below the collapse line, and thus will never undergo core
collapse. Most globular clusters, on the other hand, lie well above the
line. In such a case, the cluster may or may not go through core
collapse, depending on its initial size. We show how an accurate age
determination can help settle this issue.
\end{abstract}


\section{Introduction}
\label{sec:intro}
In the classic theory of cluster evolution, the interior region contracts
and transfers energy through distant two-body encounters to an
expanding outer halo. This process of relaxation is a consequence of
the negative heat capacity of self-gravitating systems. Eventually,
the rise of central density becomes dramatic, an event known as the
gravothermal catastrophe \citep{1968MNRAS.138..495L}.  The system at
this epoch is commonly said to undergo core collapse.\footnote{In this
  paper, we also adopt this terminology.  However, as we discuss
  later, our usage is more restricted than the currently popular one.}

For a hypothetical cluster comprised of identical, single stars,
numerical simulations find that core collapse occurs after more than ten relaxation times \citep{cohn1980, makino1996}. The catastrophe ends when a tight
binary forms near the center.
 Binary heating then not only halts the
rise in central density, but leads to a global expansion of the system \citep{1965AnAp...28...62H,1971Ap&SS..13..324A}. Observationally, a substantial fraction of the most massive
globular clusters ($M_{\rm cl} \gtrsim 10^5\,\msun$)  exhibit a central peak in
stellar density and luminosity, and are thought to have undergone core collapse in the
past \citep[][]{1986ApJ...305L..61D,1995AJ....109..218T,1997A&ARv...8....1M}.

Of course, the member stars of real clusters are not
identical. Globular clusters are so old that stellar evolution has
pared down the initial mass distribution to a relatively narrow
range. In this case, the classic theory provides a reasonably accurate
description at sufficiently late times.
 The vast majority of clusters in
the Milky Way are open clusters, which are much less populous and do not live nearly as long as
globular clusters \citep{2010AN....331..519R}. The evolutionary path of these more modest systems
may be  strikingly different.

Regardless of the cluster's precise initial state, its most massive
stars quickly sink to the center through dynamical friction. This mass segregation occurs in less than a single relaxation
time \citep[e.g.,][and references therein]{2004ApJ...604..632G}. Three-body encounters between the central stars soon
create a few massive binaries, whose heating frustrates the process of
core contraction \citep{2011MNRAS.410.2787C,2012NewA...17..272T}. In systems of
relatively low population and size, core collapse never occurs.

In this paper, we explore these two, very different evolutionary
paths. We first delineate the boundary between the two paths using an
analytic argument. We then verify the location of that boundary and
describe the structural evolution of clusters on either side of it
numerically. Here, we employ the direct $N$-body integrator {\sc
  nody6-gpu} \citep{nbody6,GPU}. Our simulations span a population
range from about $8\times10^3$ to over $1 \times 10^5$ so that we may
characterize the evolution of both open and globular clusters. Our
simulations include a realistic mass distribution, stellar evolution,
and the influence of the Galactic tidal field.

Our combined analytical and numerical results show that a cluster must
be relatively large and populous to undergo core collapse. Because of
the external tidal field, the evolution of any system also depends on
its location relative to the Galactic center. For any fixed location,
there exists a minimum population such that sparser clusters never
undergo collapse, regardless of their exact initial state. We again
verify the existence of this minimum population both analytically and
numerically.

A number of previous studies also investigated cluster evolution
through a suite of simulations. In some cases, the
researchers included populations as high as our maximum
value \citep[e.g.,][]{2003MNRAS.340..227B,2011MNRAS.411.1989Z}. 
These projects addressed a variety of specific issues, such as how
metallicity effects the changing appearance of clusters \citep{2012MNRAS.427..167S}.  \citet{2003MNRAS.340..227B}
presented the most extensive, simulation-based investigation to date,
focusing primarily on how the stellar mass function
scales with time. Ours is a complementary study, intended to establish the broad landscape in which
clusters evolve.

This work is organized as follows. In \S~\ref{sec:time}, we analyze
the relevant timescales for the competition between mass segregation,
relaxation, and stellar mass loss in clusters. We describe our
numerical simulations in \S~\ref{sec:simulations} and present the bulk
of our results in \S~\ref{sec:results}. Finally, we summarize and discuss the implications of our results in \S~\ref{sec:disc}.

\section{The collapse line}
\label{sec:time}
We focus here on the cluster's evolution after the first few dynamical
times since its formation within a molecular cloud. By that point,
radiation pressure and energetic winds from the most massive stars
have dispersed all cloud gas, leaving the stars to interact only via
their mutual gravity. Subsequently, the bulk of the cluster steadily
expands, until the system is ultimately disrupted tidally, either by
passing molecular clouds or the Galactic field. The central issue we
address is the fate of the cluster's deep interior.

The evolution of this central core is driven by the competition of
dynamical cooling and heating \citep[see, e.g.,][]{henon1961}. On the
one hand, the core transfers energy outward to the halo stars through
two-body relaxation, and thereby tends to contract. On the other hand, a
single hard binary near the center of the cluster may effectively heat
the core through three-body interactions, causing it to expand.  The
core contains a large fraction of the cluster's most massive stars.
In the course of stellar evolution, mass loss from these objects in
the form of stellar winds and supernovae diminishes the gravitational
binding of the core, again promoting expansion.

Consider first a hypothetical cluster of mass $M_{\rm cl}$ containing $N$
identical stars. Here, the central core transfers energy outward
through two-body encounters. This transfer occurs on $t_{\rm rel}$, the initial relaxation time scale
\begin{equation}
\label{eq:relax}
t_{\rm rel} = \frac{0.17 N}{\ln(0.1 N)}\sqrt{\frac{r_v^3}{G M_{cl}}},
\end{equation}
 which we define at the virial radius, $r_v$  \citep[cf.][eq. 7.108]{2008gady.book.....B}. We take the virial radius to be $r_v \equiv G M_{cl} / 6 \sigma^2$, where $\sigma$ is cluster's one-dimensional velocity dispersion.  When tracking $r_v$ in
our simulations, we determined $\sigma$ directly from the stellar
velocities at each time step.

While the uniform-mass cluster can, in
principle, form binaries, the time to do so is several hundred $t_{\rm rel}$ \citep[][eq. 7.12]{2008gady.book.....B}. Long before this, the cluster's central density rises,
eventually in a divergent manner in a finite time. Numerical simulations suggest that
this core collapse occurs at  $t_{cc} \approx 16
t_{\rm rel}$ \citep{cohn1980,makino1996}. Despite the idealized
assumption underlying this picture, it is still frequently used as the framework to
describe the evolution of all clusters.

Realistic clusters have a broad spectrum of stellar masses. The most
massive stars, whatever their initial location, migrate toward the
center as a result of dynamical friction. For a star with mass $m_* \gg \mean{m}$,
 this
process occurs over the dynamical friction time  $t_{df}$, which is brief
compared to $t_{\rm rel}$:
\begin{equation}
\label{eq:tdf}
t_{df} \approx \frac{\mean{m}}{m_*} t_{\rm rel}.
\end{equation}
Here, $\mean{m}$ is the cluster's average stellar mass \citep{2002ApJ...570..171F}.  Both the mass
density and average stellar mass in the core are thereby enhanced.  

Since the relaxation time in the core is smaller than that of the
cluster as a whole, the core effectively decouples from the rest of
the system, and evolves separately, a process known as the \citet{1969ApJ...158L.139S} instability. For decoupling to
occur, a sufficient number of massive stars must migrate to the
cluster center, forming a subgroup whose dynamical temperature, proportional to the mean value of $m \sigma^2$, rises above that of surrounding stars. The time required for
decoupling, $t_{\rm dec}$, depends on the density profile
of the cluster and the degree of primordial mass segregation \citep[e.g.,][]{1996NewA....1..255Q,2009ApJ...698..615V}.
 A representative value, adequate for our purpose, is $t_{\rm dec} = 2\,t_{df}$.

One significant result of the core's fast evolution is the formation
of hard binaries consisting of relatively massive stars. These
binaries are created through three-body interactions. Their formation
time is very sensitive to the largest mass involved, and scales
approximately as $m_*^{-10}$ \citep{2005MNRAS.358..572I,2011MNRAS.410.2787C}. Very soon after
decoupling, at least one massive binary forms, heats the core through
three-body interactions, and creates global expansion of the cluster.
This binary can be disrupted or ejected from the cluster; however, a
new one soon replaces it  \citep{heggiehut}.

 In a cluster of sufficiently high population, stellar evolution
prevents the formation of massive binaries in the core. Let $t_{\rm ms}$
represent the main-sequence lifetime of the most massive stars. As we
consider clusters of larger $N$ and comparable size, both $t_{\rm rel}$ and $t_{\rm dec}$
grow.  When $t_{\rm dec}$ exceeds
$t_{\rm ms}$, the most massive stars explode as supernovae before they can
migrate to the center and form binaries. It is true that the
disappearance of massive stars {\it already} located in the core
temporarily heats the cluster, but eventually, the region begins to
contract via two-body relaxation. These populous, aging clusters have
a relatively narrow mass distribution, and evolve toward core collapse
in a manner similar to that of traditional theory.

\begin{figure}

\centering \includegraphics[width=\columnwidth]{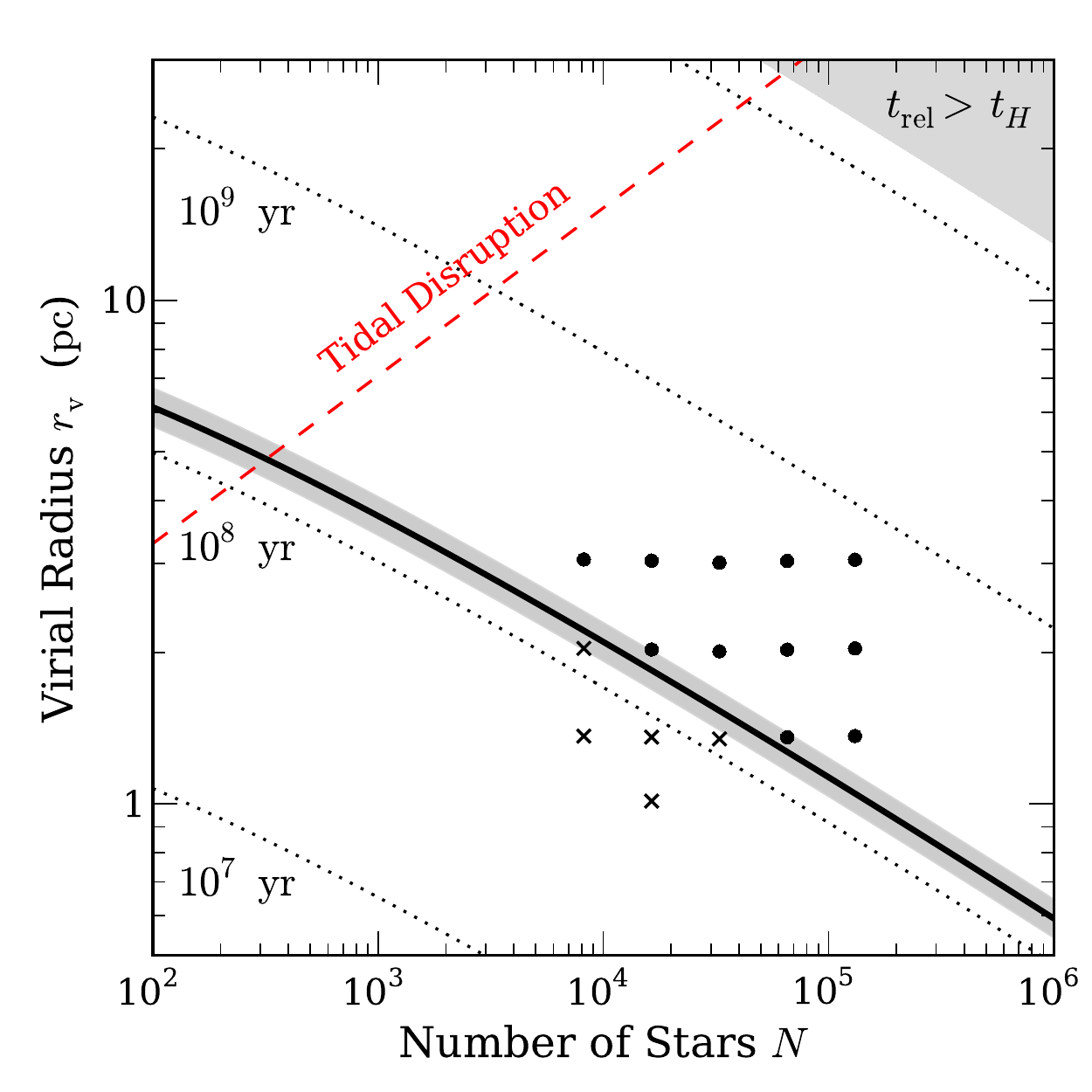}
\caption{\label{fig:init} The collapse line in the $r_v$-$N$
  plane. Clusters starting above the solid curve undergo
  core collapse if they survive long enough. Those starting below this
  curve never experience core collapse. Finally, clusters starting
  above the diagonal, dashed line are tidally disrupted at a Galactocentric distance of 8.5 kpc. Each
  dotted, diagonal line represents the indicated relaxation time
  $t_{\rm rel}$. Clusters in the upper, shaded region have $t_{\rm
    rel}$ greater than $t_{H}$, the Hubble time, and so never
  relax. The shading around the solid curve shows the range of values
  it may take depending on the clusters' precise initial conditions.
  The discrete symbols show the initial conditions of our
  simulations. Crosses represent clusters that failed to achieve core
  collapse, while filled circles are clusters whose central number
  density eventually increased.
}
\end{figure}

We see that the condition $t_{\rm dec} \approx t_{\rm ms}$ represents a dividing
line between clusters that can undergo core collapse and those that
cannot. If we elevate this condition to an exact equality, then equations~(\ref{eq:relax})~and~(\ref{eq:tdf}) may be used to solve for $r_v$ in terms of $N$, 
\begin{equation}
\label{eq:collapse}
r_v^{3/2} =    2.9\,t_{\rm ms}  \sqrt{G \mean{m}} \frac{m_*}{ \mean{m}} \frac{\ln(0.1 N)}{\sqrt{N}}.
\end{equation}
In this equation $m_*$ is the upper limit to the stellar mass.  
Any star with $m_* \gtrsim 10\msun$ has a $t_{\rm ms}$ shorter than
$2\times 10^7$ yr, a time approaching that during which the cluster
was still embedded in its parent molecular cloud.\footnote{ Optically
  visible clusters younger than 10 Myr do exist, but are relatively
  rare.  In the catalog of 642 open clusters published by
  \citet{2012A&A...543A.156K}, only 26 have tabulated ages less than
  10 Myr.} 
 If we choose $m_* = 10\,\msun$ and a minimum mass of
$0.1\,\msun$, then the average mass is $\mean{m} \approx 0.6\,\msun$
for the initial mass function of \citet{scalo}.  Setting $t_{ms} =
2\times 10^7$ yr we find the {\em collapse line}:
\begin{equation}
\label{eq:criteval}
r_{\rm coll} = 2 \left(\frac{6.9+\ln{N_4}}{6.9 \sqrt{N_4}} \right)^{2/3}{\rm pc},
\end{equation}
where  $N_4 \equiv N / 10^4$, and $6.9 = \ln{(10^3)}$.
We plot equation~(\ref{eq:criteval}) as the solid line in Figure~\ref{fig:init}. The dotted, diagonal lines in the figure represent the indicated
values of $t_{\rm rel}$. Systems within the shaded region in the upper right corner can never
relax, since $t_{\rm rel}$ exceeds the Hubble time, $t_H = 14\,$Gyr.

In numerically evaluating the right side of equation~(\ref{eq:collapse}), we have made certain definite, but somewhat arbitrary, choices.
 For example,
the literature offers a number of prescriptions for the field-star
initial mass function. Using the \citet{2003ApJ...598.1076K} initial mass function with the  same upper and lower bounds for the stellar mass would  reduce $\mean{m}$ to $0.45\,\msun$
and thus shift the collapse line in Figure~\ref{fig:init}
to the right by $\Delta \log{N} \approx 0.12$, while preserving its slope. Retaining the Miller \& Scalo initial mass function, but increasing $m_*$
by a factor of two would lower $t_{\rm ms}$ to $1\times10^7\,$yr, still a reasonable
estimate for the embedded duration. Since the product
$(t_{\rm ms} m_*)$ is insensitive to $m_*$ in this regime, the collapse line
would shift only slightly to the left.   Primordial mass segregation and variations in the initial density profile of the cluster introduce a similar amount of uncertainty. It is therefore more accurate
to envision the collapse line we plotted in Figure~\ref{fig:init} as a narrow band of total width  $\Delta \log{N} \approx 2\log{\Delta \mean{m}} \approx 0.2$. We indicate this band by
light shading in the figure.

\section{$N$-body Simulations}
\label{sec:simulations}

In this work, we have performed an extensive suite of $N$-body
simulations using the direct $N$-body integrator {\sc nody6-gpu}
\citep{nbody6,GPU} accelerated with graphics processing
units.\footnote{\url{http://www.ast.cam.ac.uk/~sverre/web/pages/nbody.htm}}
Our simulations include both single and binary star evolution,
treating mass loss as an instantaneous process
\citep{2000MNRAS.315..543H,2002MNRAS.329..897H}.  We also include a
representation of the Galactic tide, as described below.  We focus
on the evolution of star clusters after the primordial gas has been
removed from the system by both low-mass stellar outflows and by the
ionization and winds from massive stars.

We initialize our clusters with single stars distributed in a
\citet{plummer} potential,
\begin{equation}
\label{eq:plummer}
\phi_{\rm pl}(r) =  -\frac{G M_{\rm cl}}{\sqrt{r^2+a^2}},
\end{equation}
where $a = (3 \pi / 16) r_v \approx 0.59\,r_v$ is the Plummer radius and $M_{cl} = N \mean{m}$.
The mass of each star is then selected between $0.1\,\msun$ and
$10\,\msun$ following the initial mass function of
\citet{scalo}.  After generating the stellar distribution, the masses are rescaled so that the maximum stellar mass is exactly $10\,\msun$, with $\mean{m} \approx 0.6\,\msun$.  For simulations with a fixed $N$, both the masses
of all stars and their positions scaled to the virial radius are
identical, to reduce the stochastic noise.
 No initial mass segregation is imposed. 

The
early evolution of a cluster depends in detail on the fraction and
spatial distribution of primordial binaries \citep[see, e.g.,][]{2007MNRAS.374...95P,2007MNRAS.374..344T}. Rather
than explore this dependence, we have chosen to start all runs with
single stars only. This choice gives us a
uniform set of initial conditions and, in any case, has little
practical effect on the subsequent evolution. As has been shown in 
previous studies \citep[e.g.,][]{2012NewA...17..272T}, and as we verify, central binaries rapidly
form after the more massive cluster members drift to the center via
dynamical friction.

We select the cluster virial radii $r_v$ and sizes $N$  to cover the transition in evolutionary path from global expansion to
core collapse for systems similar
to those observed in the Milky Way. In Figure~\ref{fig:init}, we show
the initial conditions for 16 of our simulations which bracket the
collapse line (eq.~\ref{eq:criteval}).
The clusters have populations $N$ starting from
8,192 and increasing, by a factor of two, to a maximum of $N = $131,072.
The minimum $r_v$-value, used only in conjunction with $N = $16,384, was
1.0 pc. For all other $N$-values, we used $r_v$ = 1.3, 2.0, and 3.0 pc.
The upper limit for $N$ was chosen for practical reasons.
 Simulations with $N = 131,072$ took a few weeks to complete
on a single desktop with a GPU. Increasing the cluster population even by a factor of two would have
required several months per run.

The clusters evolve in a spherical tidal gravitational field similar to that
experienced by Milky Way clusters on circular orbits. Specifically, we
adopt an isothermal potential with a constant circular velocity, $v_c
= 220\,\kms$. With this potential, there is an enclosed mass of ${M_{\rm enc}=9.6\times10^{10}(R_G/8.5\,{\rm kpc})\,\msun}$ within an orbit of radius
$R_G$. Unless otherwise noted, all simulations assume the clusters
follow a circular orbit of radius $8.5\,$kpc.
Stars are removed from our simulations when they are outside of $2r_t$, where $r_t$ is the 
 tidal radius:
\begin{equation}
\label{eq:rt}
r_t \equiv R_G (M_{\rm cl} / 2 M_{\rm enc})^{1/3}.
\end{equation}
For our models, the value of $r_t$ ranges from $20$ to $60\,$pc.

We run all of our simulations for a minimum of $15\,t_{\rm rel}$. If
the cluster does not exhibit core collapse by $t=15\,t_{\rm rel}$, we
continue the simulations until the system loses at least $90\%$ of its initial population. In
none of these cases did core collapse occur before the cluster
dissolved.

\section{Results \& Analysis}
\label{sec:results}

As first envisioned, core collapse occurs when the
central density of a cluster rises in a sharply accelerating manner.
Such behavior was first predicted theoretically \citep{henon1961,1968MNRAS.138..495L}, then clearly
exhibited in both fluid models of clusters \citep{1970MNRAS.147..323L} and in $N$-body
simulations of idealized systems comprised of identical-mass stars
\citep[e.g.,][]{1974A&A....37..183A,makino1996}. In these cases, there is no ambiguity in defining the central
density or describing its temporal change. However, subtleties arise
when analyzing modern simulations that follow the dynamics of a
stellar population spanning a realistic distribution of masses.

Our main goal is to give an account of cluster evolution that will
prove useful when considering real, observed systems. These are only
seen in projection against the plane of the sky. Hence, we begin by
discussing the evolution of the projected, two-dimensional central
number density, $\Nc$. We focus on the number,
rather than mass density, since the latter may change because of
local processes, such as stellar mass loss via winds. We defer
discussion of three-dimensional effects to the following subsection.
There we also view our results within an alternative framework that is
also frequently employed --- the evolution of the cluster's core radius.

\subsection{Evolution of the projected central density}
\label{sec:2d}

Before even choosing an operationally suitable definition of $\Nc$, we
must first be able to locate with precision the cluster's center.
Following \citet{1960ZA.....50..184V}, we first associate a local surface density ${\cal N}_i$
with each cluster member, excluding escapers, here labeled by the index $i$. This member can
be a single star or a binary. To establish each ${\cal N}_i$, we use the area
spanned by the object's 7 nearest neighbors \citep{core}.
Then, relative to any convenient origin, the cluster center is the
density-weighted average position vector of all the members:
\begin{equation}
\label{eq:center}
{\bm{R}_{\rm o}} = \frac{\sum_i  {\bm R_i} {\cal N}_i}{\sum_i {\cal N}_i},
\end{equation}
For cluster members that are binaries, $ {\bm R_i}$ locates the center of mass
of the pair. Note again that all vectors are two-dimensional.

\begin{figure*}
\centering \includegraphics[width=\textwidth]{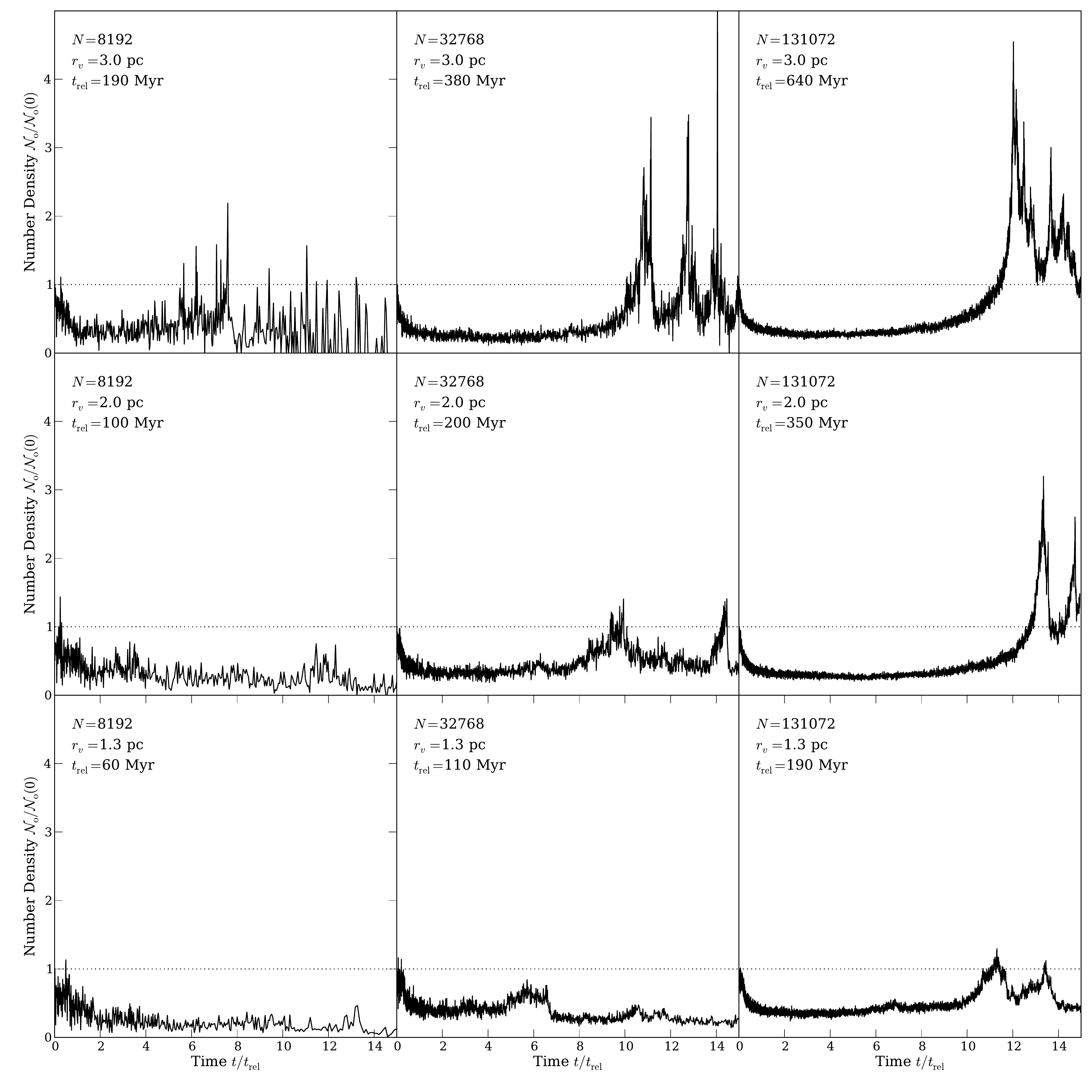}
\caption{\label{fig:3x3} Normalized central surface number density. We
  plot the central projected number density of stars  within $0.01 R_h$, $\Nc$,
  normalized by the initial density, $\Nc(0)$, as a function of time for
  nine of our simulations. The number of stars, $N$, and virial
  radius, $r_v$, of each cluster is labeled in the panel. Also labeled is the initial relaxation time, $t_{\rm rel}$. The dashed
  line shows the threshold for core collapse. In general, the depth of
  the collapse increases with $N$ (from left to right) and $r_V$ (from
  bottom to the top). Only the clusters above the collapse
  line (see Fig.~\ref{fig:init}), however, undergo core collapse.}
\end{figure*}

\begin{figure*}
\includegraphics[width=\tw,type=pdf,ext=.pdf,read=.pdf]{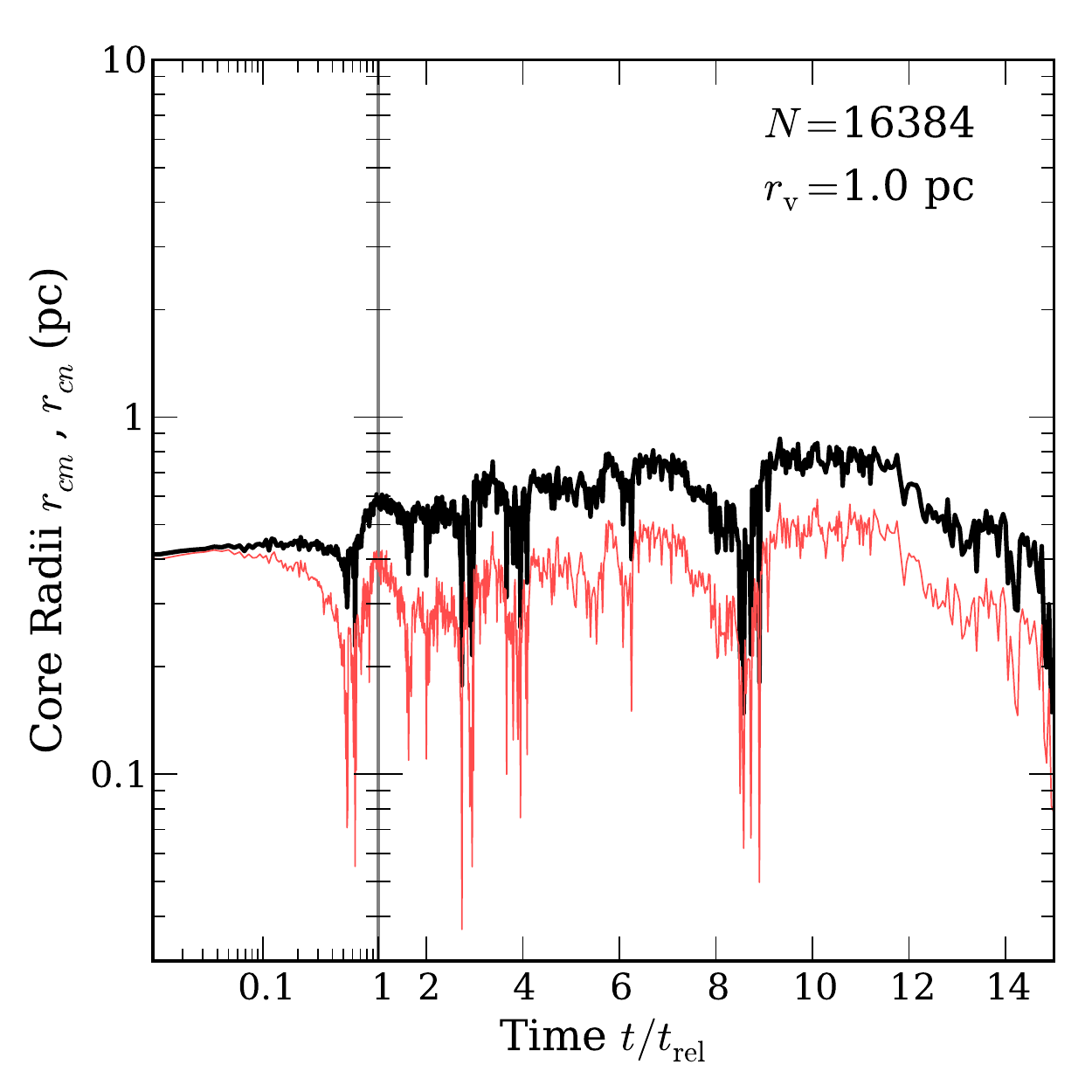}
\includegraphics[width=\tw,type=pdf,ext=.pdf,read=.pdf]{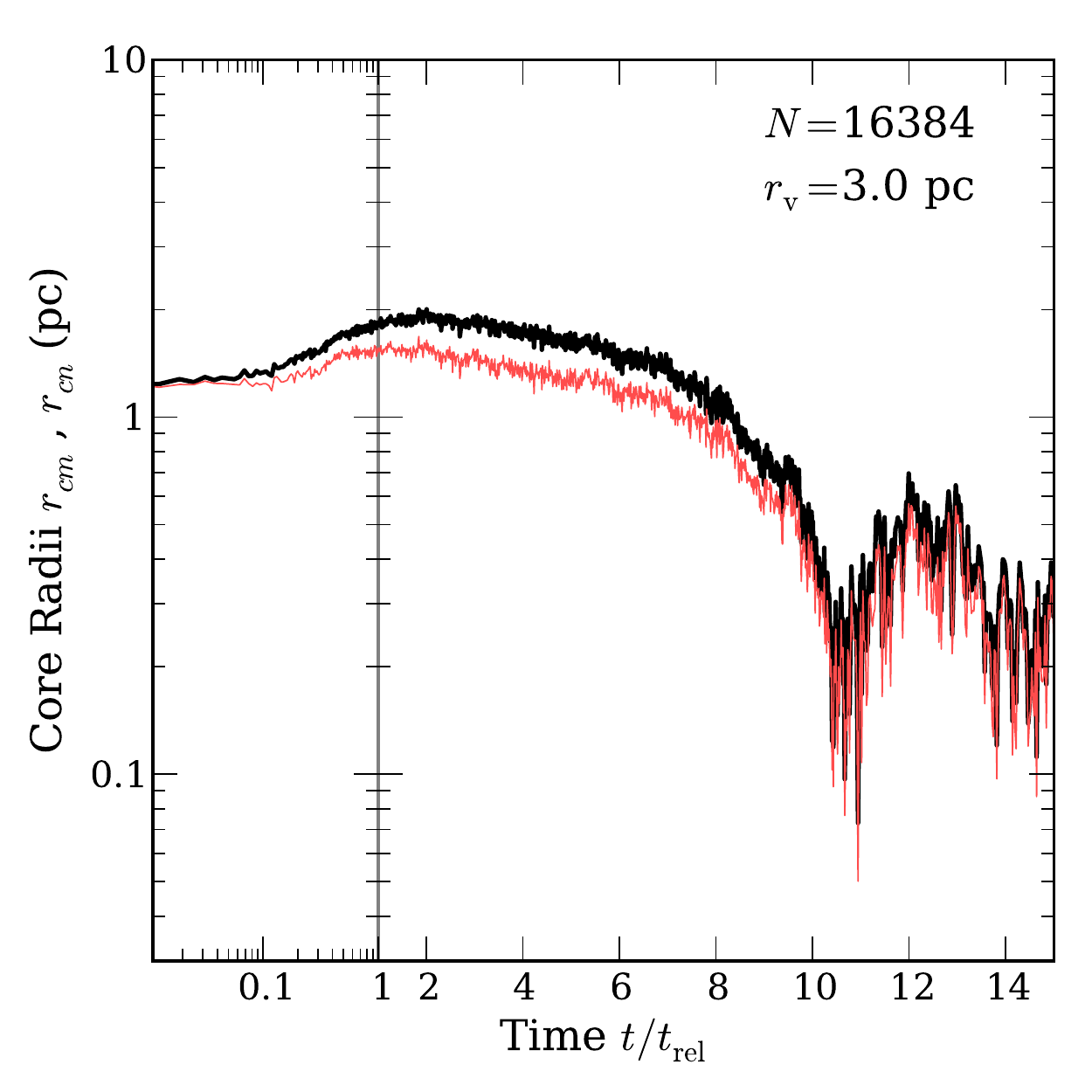}

\caption{\label{fig:core}  Evolution of the core radii $r_{cn}$ and $r_{cm}$ for a
cluster that does not undergo core collapse ({\it left panel}) and one
that does ({\it right panel}). In both cases, the number-weighted
radius $r_{cn}$ is the thicker curve that is higher in the plot. As in
Figure~\ref{fig:3x3}, we measure time relative to the initial value of $t_{\rm rel}$.  Note the scale of the  horizontal axis changes from logarithmic to linear at $t/t_{\rm rel} = 1$.}
\end{figure*}

Having located the center, we find $\Nc$ by counting up the cluster
members within a concentric circle. If the circle is chosen to enclose
a fixed and relatively small number of stars, eg, 7 or 70,  then the defined $\Nc$
undergoes large Poisson fluctuations. In addition, the adopted radius should
scale with the total cluster population $N$, which varies widely in our
suite of simulations. Let $R_h$ be the radius
containing, at any time, half the total members. If $f$ is a number less
than unity, then the central density $\Nc$ is defined as the number of
stars within the radius $f R_h$, divided by the corresponding surface
area. Choosing an $f$-value that is relatively high (e.g., 0.1) smears out
the evolution. We have found in practice that $f = 0.01$ yields an $\Nc$
that exhibits a smooth evolution while retaining easily discernible
trends. Obviously, there is some latitude in this definition; changing
$f$ by a factor of two or so in either direction makes no appreciable
difference in the resulting $\Nc (t)$, except for the amount of noise in the results.

 To undergo core collapse, the cluster should exhibit an accelerating
 rise in central density. In particular, we {\em define} core collapse
 to occur when the central density of the cluster exceeds its initial
 density\footnote{To minimize Poisson fluctuations, we determine the
   initial central density by taking the maximum value over the first
   20 dynamical times. If we instead took the average value, the cluster would, according to our criterion, start in a core-collapsed state.}, 
\begin{equation}
\label{eq:ccdef}
\frac{\Nc}{\Nc(0)} > 1.
\end{equation}
  In Figure~\ref{fig:3x3}, we plot $\Nc$ as a function of time for
  nine runs.  Here we have normalized $\Nc$ by $\Nc(0)$, and the
  time $t$ by the initial value of $t_{\rm rel}$.  All clusters with an initial
$r_v$ of $3.0\,$pc ({\it top row}) undergo
core collapse, according to our criterion. Of clusters starting with
$r_v = 2.0\,$pc ({\it middle row}), those two with $N \ge 32768$ undergo
collapse. Finally, of the clusters with $r_v = 1.3\,$pc ({\it bottom
row}), only that with the highest $N$ reaches this state. In all these
cases, the initial parameters of the
  clusters place them above the collapse line in
  Figure~\ref{fig:init}.  These systems have relatively large $N$ and
  most closely mimic the behavior of uniform-mass clusters.

In all such runs, we see that the central density, after exceeding its
initial value, later plunges below it and then climbs again. If the
simulation were extended to longer times, this pattern would repeat.
Such ``gravothermal oscillations'' are caused by the successive
formation and ejection of central binaries. The phenomenon has long
been documented in the theoretical literature on equal-mass systems \citep{1984MNRAS.208..493B, 1984ApJ...280..298G}.

In contrast to this behavior is that of clusters with $t_{\rm dec} <
t_{\rm ms}$, i.e., those starting below the collapse line in
Figure~\ref{fig:init}. In these, the central density never exceeds its
initial value, except perhaps transiently early in the
evolution. (Such excursions last about one dynamical time.) These
systems undergo global expansion, as was found in the simulations of
\cite{2011MNRAS.410.2787C}, and as we will show in more detail
below. In Figure~\ref{fig:init}, we have marked the model clusters
that undergo core collapse with filled circles, and those
exhibiting global expansion with crosses. It is evident that the
collapse line indeed demarcates the two distinct evolutionary paths.

The maximum normalized central density attained by a cluster increases
with both $N$ and $r_v$, from the bottom left to the top right in Figure~\ref{fig:3x3}. Graphically, the systems that reach the highest central density
during core collapse are farthest above the collapse line. For systems
lying close to the line, relatively small changes in $N$ or $r_v$ can
result in the central density either falling a bit below or slightly
above its initial value. In such marginal cases, it is unclear whether
the cluster should be deemed as undergoing core collapse. Further, we
have noted that our operational definition of the central density
itself is somewhat arbitrary. These factors introduce 
additional uncertainty in the true location of the collapse line, but
the induced width is small compared to that arising from initial
conditions, as outlined in Section~\ref{sec:time}.

\subsection{Three-dimensional evolution: Core radius}
\label{sec:3d}

We gain a better physical understanding of the cluster's behavior by
examining it not just in projection, but in three-dimensional space.
The extensive literature in this field has focused traditionally on
the evolution of the core radius, again defined three-dimensionally.
It is instructive to view our main result in this perspective, both to
place it in the context of previous research, and to further elucidate
the two evolutionary paths.

As in Section~\ref{sec:2d}, we must first establish the cluster center, this
time in three dimensions. We again follow 
\citet{1960ZA.....50..184V} and \cite{core}, using the 7 nearest neighbors to assign a
local, volumetric number density $n_i$ to each cluster member.  We identify the
cluster's center as the density-weighted average position vector of
all members:
\begin{equation}
\label{eq:center3d}
{\bm r}_{{\rm o} n} = \frac{\sum_i  {\bm r_i} {n}_i}{\sum_i {n}_i}.
\end{equation}
Note that the additional ``$n$'' subscript specifies our use of {\it
number} densities in the weighting. We define a number-weighted core
radius, denoted $r_{cn}$, by finding the average distance of members
from the center, again weighted by the local number density:
\begin{equation}
\label{eq:core3d}
r_{cn} = \frac{\sum_i |{\bm r_i - \bm r_{{\rm o}n}}| {n}_i}{\sum_i {n}_i}.
\end{equation}

\begin{figure*}
\centering \includegraphics[width=\tw,type=pdf,ext=.pdf,read=.pdf]{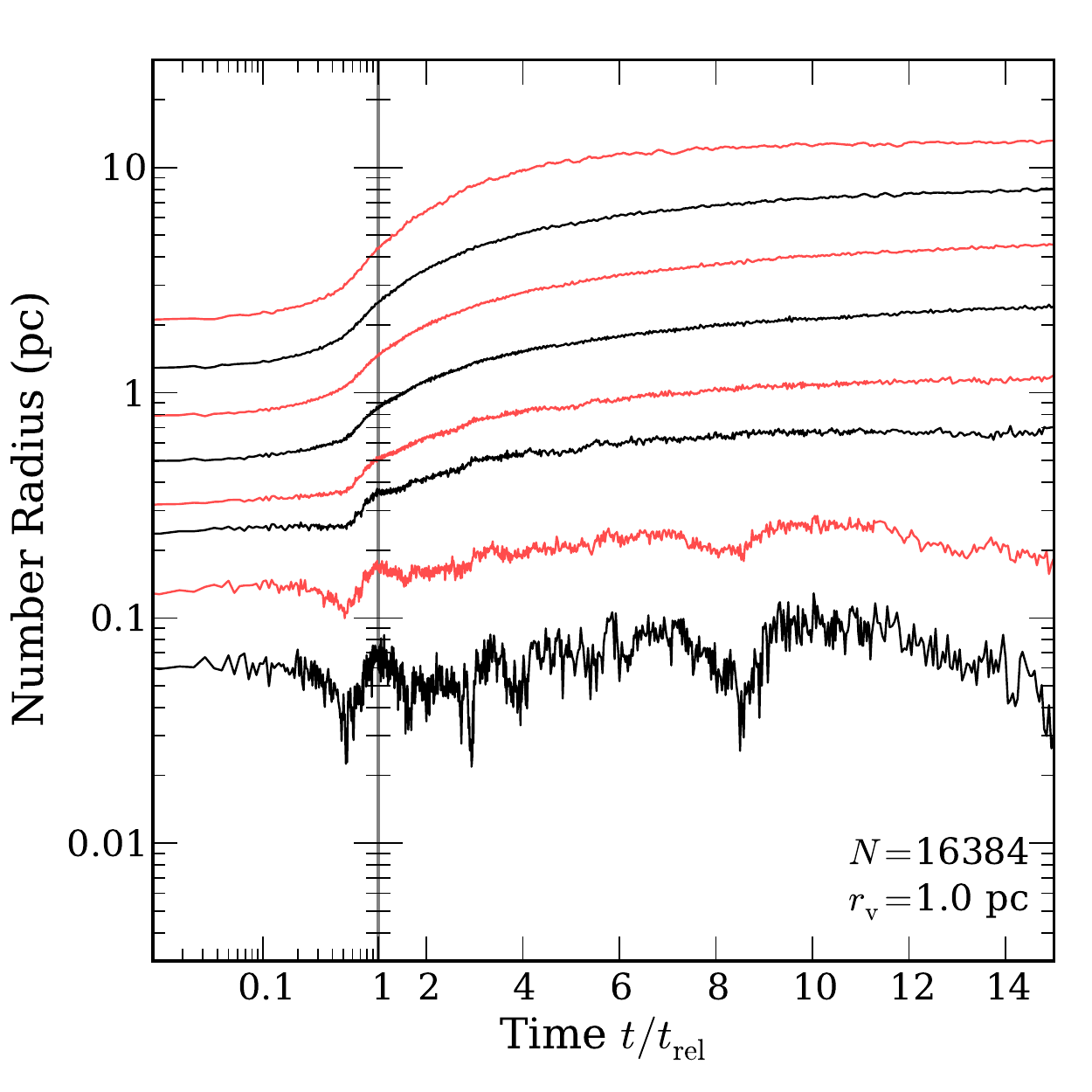}
\includegraphics[width=\tw,type=pdf,ext=.pdf,read=.pdf]{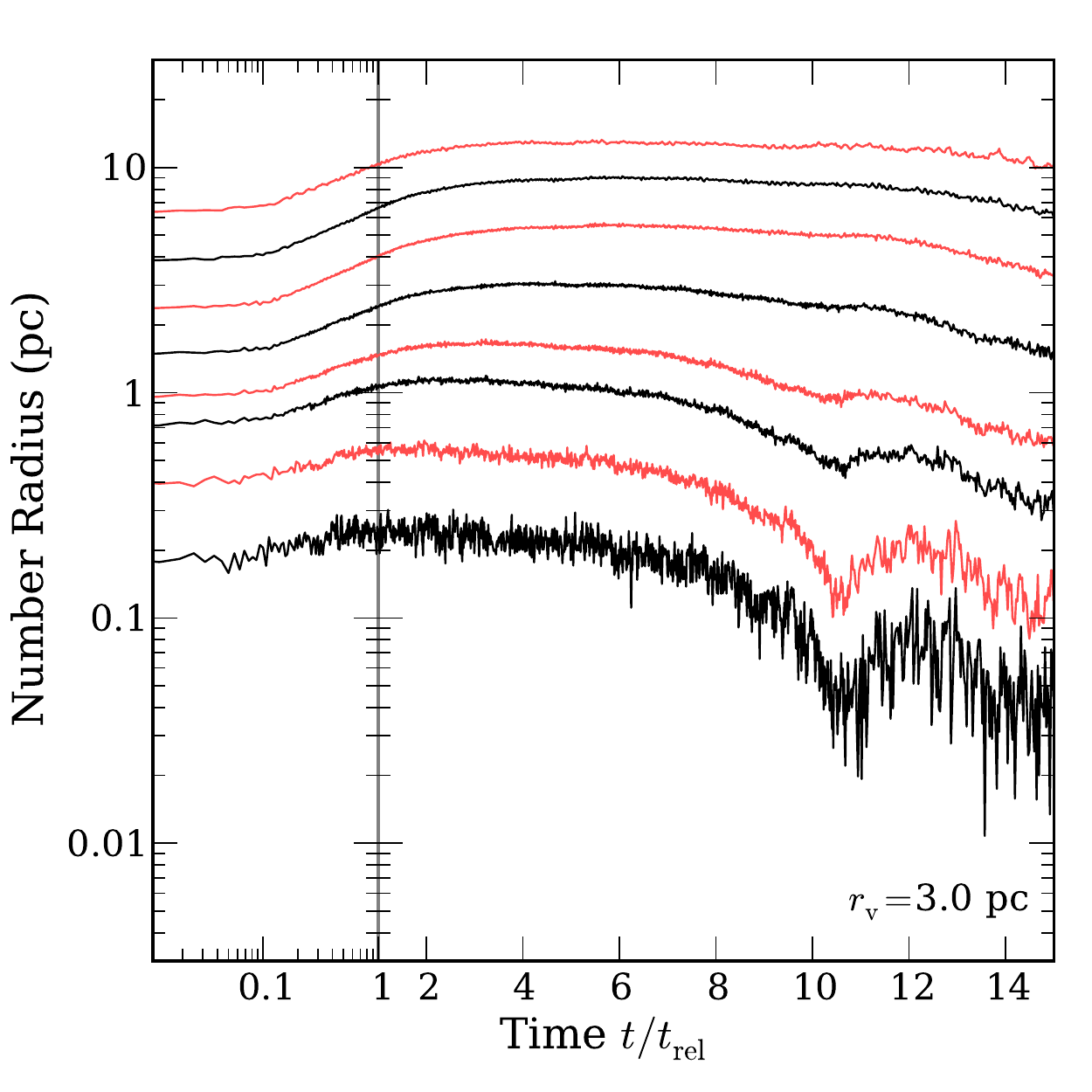}
\\
 \includegraphics[width=\tw,type=pdf,ext=.pdf,read=.pdf]{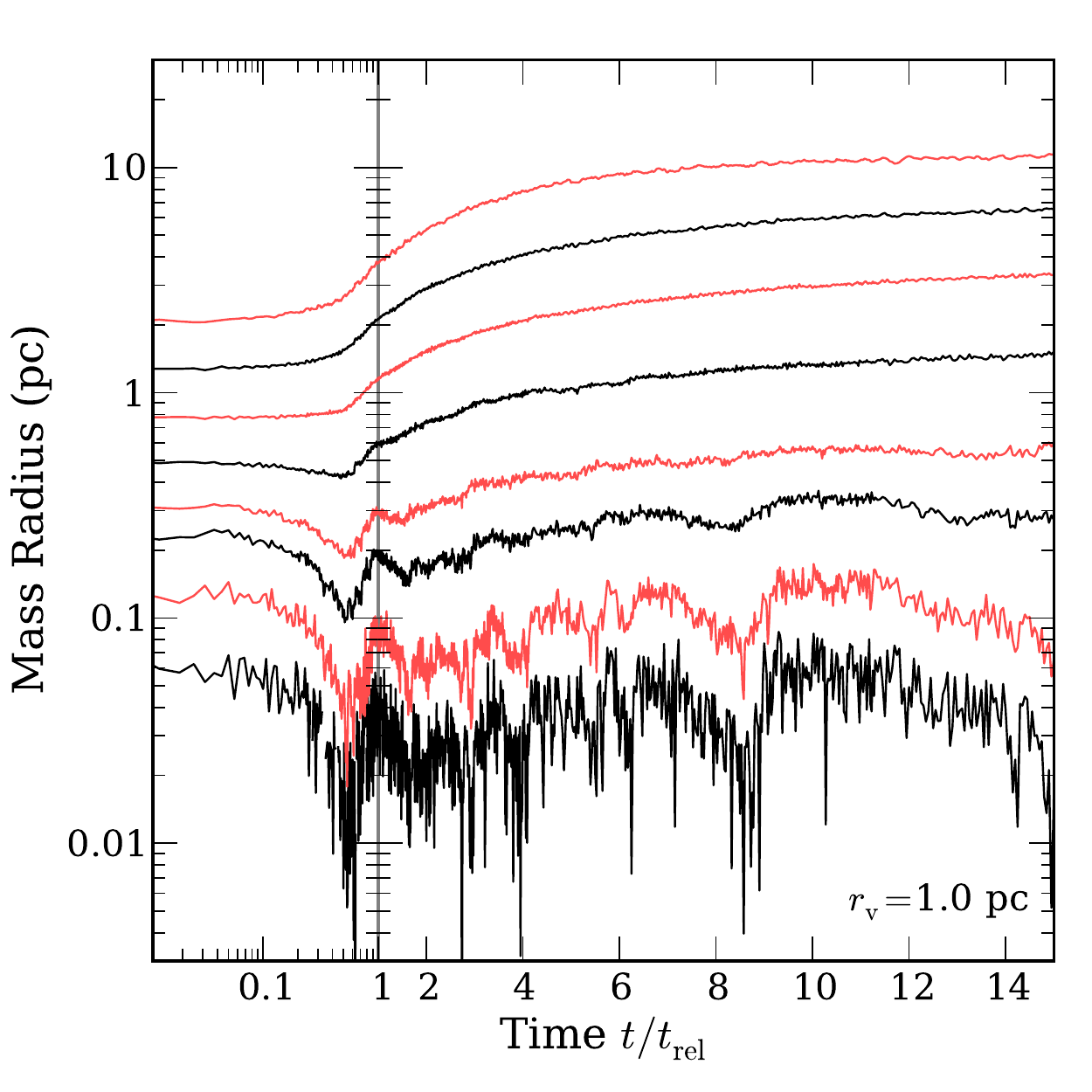}
\includegraphics[width=\tw,type=pdf,ext=.pdf,read=.pdf]{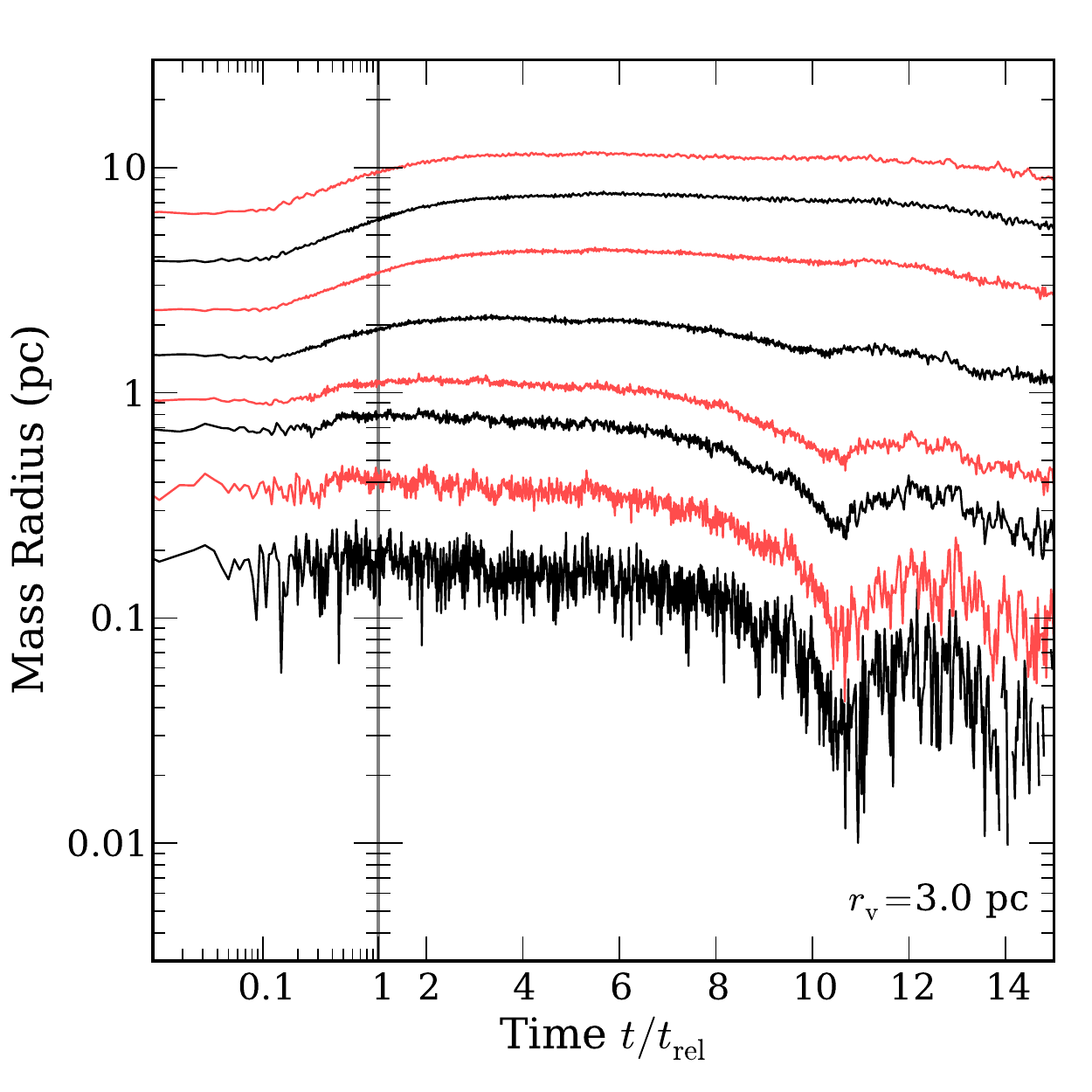}
\caption{ \label{fig:lagr} Evolution of the number radii ({\it top row}) and mass radii ({\it bottom row}) for a cluster with $N = $16,384. For a starting virial
radius of $r_v = 1.0\,$pc, the cluster does not undergo core collapse,
while it does for $r_v = 3.0\,$pc. Within each panel
 the displayed
  Lagrangian radii enclose a fraction $f = 0.001, 0.01, 0.05, 0.10,
  0.25, 0.50, 0.75,$ and $0.90$  of the cluster's total population
  or mass.  Note again that the scale of the horizontal axis changes from
  logarithmic to linear at $t/t_{\rm rel} = 1$, where $t_{\rm rel}$ is the initial relaxation time.}
\end{figure*}

Our definitions of the cluster center and core radius follow those in
the literature, but with one difference. Traditionally, the weighting
factor is the {\it mass} density $\rho_i$ associated with each
member.\footnote{Some authors have used $\sqrt{\sum_i' \rho_i^2
    |\bm r_i - \bm r_{{\rm o}m} |^2 /\sum_i' \rho_i^2}$ as the
  definition of the core radius. See, e.g.,
  \citet{1990ApJ...362..522M}} Using $\rho_i$ instead of $n_i$, we may
alter equations~(\ref{eq:center3d}) and ({\ref{eq:core3d}) to define the
  analogous mass-weighted central position vector, $r_{{\rm o}m}$, and
  core radius, $r_{{\rm c}m}$. It is also traditional to track the
  evolution of the cluster's tidal radius, $r_t$.  Beyond this radius, a star escapes the cluster if the system is
on a circular orbit around the Galactic center. When
\begin{equation}
\label{eq:oldcc}  
\frac{r_{cm}}{r_t} \le 10^{-2},
\end{equation}
the cluster is traditionally said to undergo core collapse (e.g.,
\citealt{makino1996}, \citealt{2004ApJ...604..632G}, and
\citealt{2007MNRAS.374...95P}).

For the evolution of an idealized, single-mass cluster, the number and
mass densities are proportional at all times. Hence 
$r_{cm} = r_{cn}$. In this case, simulations find that the cluster's
interior density eventually exhibits a power-law profile. That is $n
(r) \propto r^{-p}$, where $r$ is the distance from the center and the
exponent $p \approx 2.2$ \citep[see, e.g.,][]{cohn1980}. At this epoch, the core radius,
as defined by equation~(\ref{eq:core3d}), shrinks to zero, and the traditional criterion for
core collapse, equation~(\ref{eq:oldcc}), is satisfied.

In more realistic systems containing a range of stellar masses, the
two definitions of core radius are not equivalent. The behavior of
$r_{cn} (t)$ and $r_{cm} (t)$ is more complex and interesting than in
single-mass models.  Figure~\ref{fig:core} displays the evolution of
the two quantities for two simulations that lie on either side of the
collapse line with $N=16,384$. In the left panel (with $r_v = 1.0\,$pc), which shows a cluster that does
{\it not} undergo core collapse, $r_{cn} (t)$ exhibits fluctuations,
but does not deviate by more than a factor of a few from its value at $t = t_{\rm rel}$.\footnote{The cluster as a whole
expands after the formation of the first binary. However, $r_{cn}$ remains
roughly constant as the number of stars in the core declines (see \S~\ref{sec:3dlag}).} 

On the other
hand, $r_{cm} (t)$ does have several dramatic plunges. These are the
result of mass segregation. At each time, a few members of especially
large mass have drifted to the center via dynamical friction. These
members couple with others to form the binaries which cause the
cluster to expand, but they do not substantially increase the central
number density. The first
such event occurs in less than a single relaxation time, as can be
seen in the left portion of the panel. The stars in question
contribute a small fraction of the cluster's total mass
\citep{2004ApJ...604..632G}, and a much smaller fraction of the total
number of stars. In this simulation, fewer than ten stars are
involved in the contraction of $r_{cm}$.

The right panel of Figure~\ref{fig:core} displays the evolution of
$r_{cn}$ and $r_{cm}$ for a cluster that {\it does} undergo core
collapse, with $r_v = 3.0\,$pc. Here, there are no early plunges of $r_{cm} (t)$ associated
with mass segregation, since the most massive stars die before they
can reach the center. The two core radii now track each other
closely. In particular, both take a sharp drop at $t \approx 11 \,t_{\rm
  rel}$, which marks the epoch of the first core collapse, according to our definition. By this point, the
cluster's mass spectrum has narrowed considerably from the initial
state, just as in observed globular clusters.  The fraction of cluster
mass within $r_{cm}$ at the formation of the first binary is similar
to the simulation in the left panel.

\subsection{Three-dimensional evolution: Lagrangian radii}
\label{sec:3dlag}

We gain a more detailed understanding of the cluster's behavior by
following the evolution of its interior density. Traditionally,
researchers have considered spherical shells that contain a fixed
fraction of the system's total current mass.  Here, we will follow
this convention, and thus trace the radii of individual mass
shells. However, we are also interested in shells that contain a fixed
fraction of the current, total population. A ``number radius''
containing, e.g., 10 percent of the population, does not generally
contain 10 percent of the cluster mass. Indeed, the differing
evolutions of the number and mass radii provide further insight into
the nature of the two evolutionary paths of the cluster and the impact
of mass segregation.

Figure~\ref{fig:lagr} displays the evolution of selected number and mass radii for
the same two clusters as in Figure~\ref{fig:core}, i.e., systems lying on either
side of the collapse line. The top two panels show number radii for
both clusters. The one starting with $r_v = 1.0\,$pc lies below the
collapse line. Here, the number radii generally expand with time. It
is only the innermost shell, containing 0.001 of the current
population, that has repeated dips in its radius. These dips
 are associated with binary formation by massive
stars, as discussed in Section~\ref{sec:3d}. Note that the shell in question
contains at most 17
stars. Only on this tiny scale does a number radius ever decrease
significantly. In contrast, 95 percent of the cluster population
expand monotonically from the start, as exemplified by the shell containing 5
percent of the cluster population.

The top right panel traces the evolution of number radii for the
cluster starting with $r_v = 3.0\,$pc and lying above the collapse line.
In this case, the number radii in the deep interior evolve more
smoothly, since there are no repeated dips associated with massive
binary formation. All radii eventually contract, and the interior ones
plunge steeply at $t = 11\,t_{\rm rel}$, the time of core collapse.

We get a very different impression when we examine the mass radii for
the same systems.  The bottom two panels show mass radii for the same two clusters. For
the one with an initial $r_v$ of $1.0\,$pc, several radii have repeated,
sharp plunges, much steeper than those of the analogous number radii.
Each plunge occurs when a few massive stars drift to the center. The
first such event coincides with the formation of the first hard
binary. At this point, the interior mass shell comprising 1 percent of
the cluster mass contracts by an order of magnitude and contains only
7 stars. Just afterward, all mass shells rapidly enlarge, a
manifestation of binary heating.

Another traditional criterion for core collapse is the contraction of
interior mass shells \citep[e.g.,][]{1997MNRAS.286..709G,2004ApJ...604..632G}. We now see that mass
segregation, and not the global relaxation of the cluster, may be responsible for this
contraction. In the present case, only the innermost 50 percent of the
cluster's mass expands for the entire run. At the same time, the
physical spacing between almost all stars, except a very few near the
center, steadily increases. Thus, the system truly undergoes global
expansion. The net efflux
of stars from the central region accounts for the fact that the core
radius $r_{cn}(t)$ stays roughly constant, as noted previously.

In the bottom right panel, we see the evolution of mass shells in the
cluster that is initially larger. For this system, the mass and number
radii track each other quite closely, i.e., there is little sign of
mass segregation. Again, there are no deep plunges of shells early in
the evolution. Massive stars that drift to the center during that
epoch die out before reaching it. When contraction finally does occur,
it involves interior number {\it and} mass shells. At this point,
about 10 percent of the cluster's total population and mass
participate in the contraction. There is large-scale energy transfer
from the interior to the outside, as documented numerically by \citet{2011MNRAS.410.2787C}. Contraction again ends with central binary formation
and subsequent rapid expansion.

\subsection{Expansion and tidal disruption}
\label{sec:tidal}

In both
clusters shown in Figures~\ref{fig:core}~and~\ref{fig:lagr}, the bulk expansion just after formation of the first
binary exhibits power-law behavior. Thus, the
virial radius $r_v$ scales as $(t - t_o)^p$, where $t_o$ is the appropriate binary formation
time. \citet{1965AnAp...28...62H} showed that such homologous expansion is expected
whenever the cluster has a steady, central heat source. He further
showed that $p = 2/3$ under these circumstances, regardless of the
detailed physical origin of the heating.

Figure~\ref{fig:expand} demonstrates that this power-law expansion is quite general, in agreement with previous studies \citep[e.g.,][and references therein]{2010MNRAS.408L..16G}.
Here we display, in a log-log plot, the evolution of $r_v$ in three
clusters with $N = $32,768.
The cluster that has an initial $r_v$ of $1.3\,$pc does not undergo core
collapse. In this case, we find that $p = 0.4$, less than the
prediction of H{\'e}non. Successively larger clusters have shallower
slopes: $p = 0.3$ for $r_v = 2.0\,$pc, and
$p = 0.2$ for $r_v = 3.0\,$pc. Expansion in the last case is largely due to
continual mass loss via stellar evolution, a process that is not
centrally concentrated. By the time of core collapse at $t = 11\,t_{\rm rel}$,
tidal stresses have begun to decrease $r_v$ drastically.

\begin{figure}
\centering \includegraphics[width=\columnwidth]{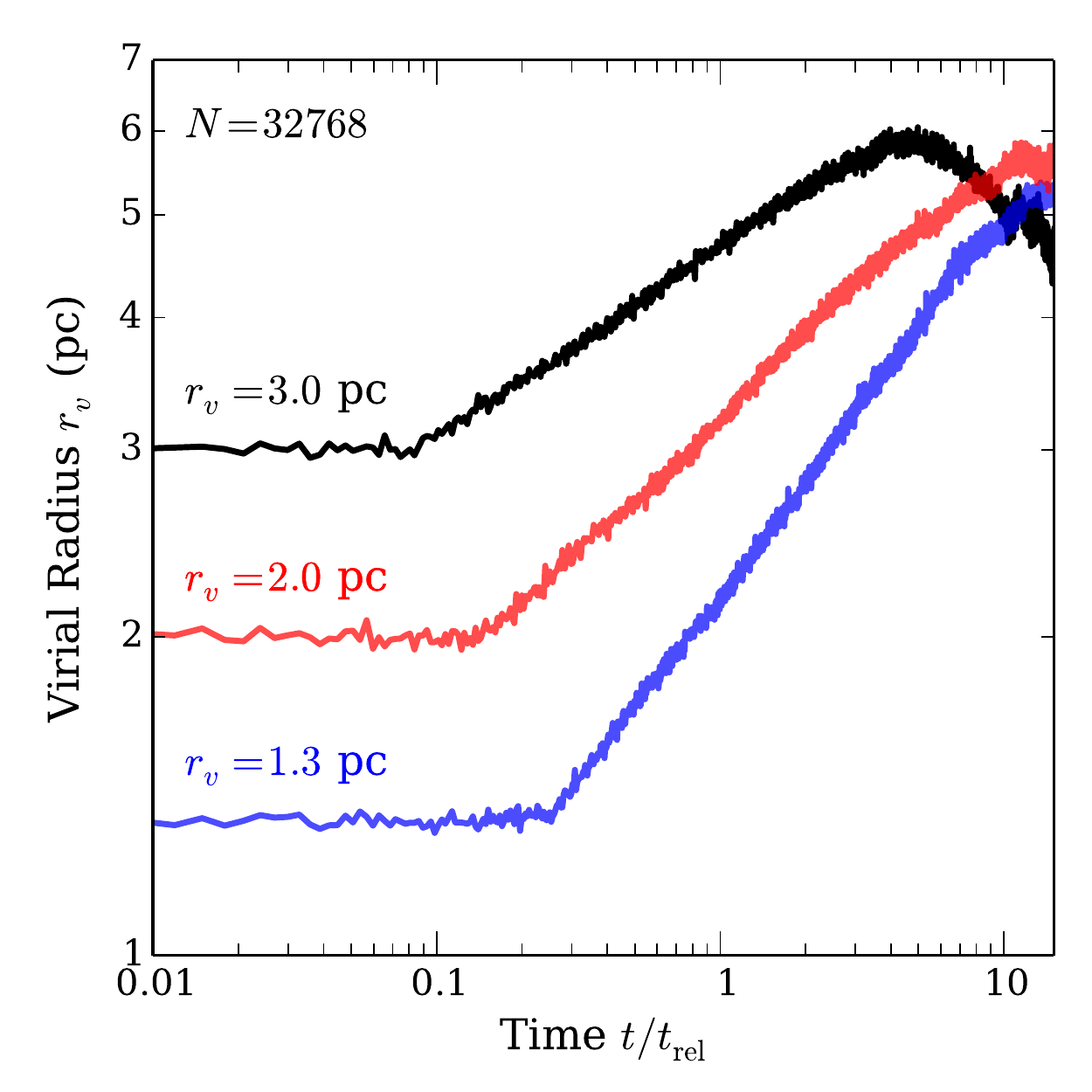}
\caption{\label{fig:expand}  Evolution of the virial radii for three clusters with $N=$32,768. Each curve is labeled by the starting values of the virial
radius. In all cases, there is an extended period of
power-law expansion, where the expansion is shallowest for the cluster with $r_v=3.0$pc. As before, $t_{\rm rel}$ refers to the initial relaxation time. }
\end{figure}

\begin{figure}
\centering \includegraphics[width=\columnwidth]{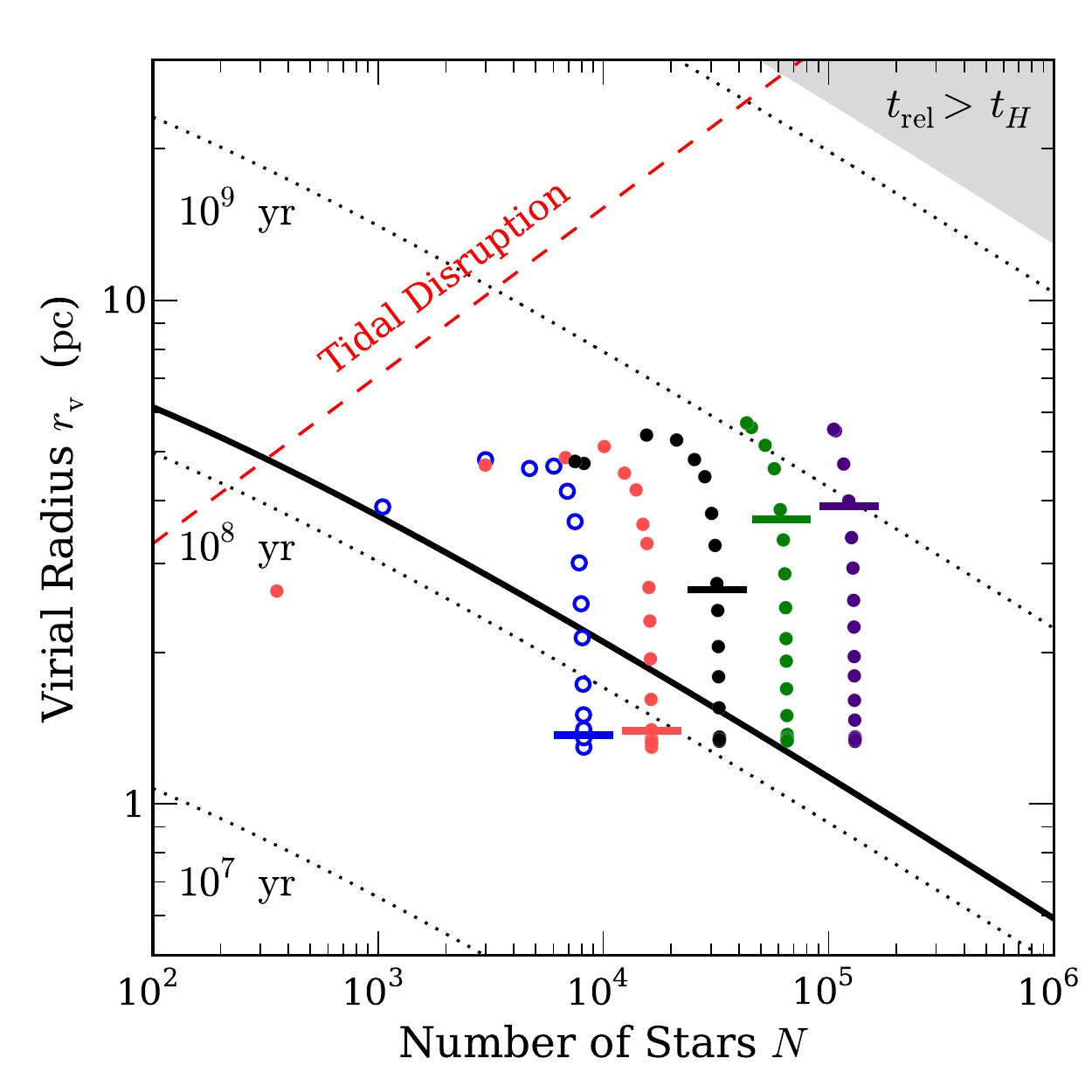}
\caption{\label{fig:tidal} 
Evolution of five representative clusters in the
$r_v$-$N$ plane. For each cluster, the horizontal lines marks where the clusters begin sustained binary formation and burning.  
All
clusters start with ${r_v = 1.3\,}$pc. For the leftmost clusters with an
initial $N$ of 32,768 or fewer, we evolve the system until disruption. We evolve
the more massive clusters  only until $t = 15 t_{\rm rel}$, before they lose half of their initial cluster members.  Note that
all curves eventually veer in a direction nearly parallel to, but below, the
tidal disruption line.}
\end{figure}

Whether a cluster's expansion is powered by central binary heating or
pervasive, internal mass loss, the distended system eventually feels
the effect of the Galactic tidal field.\footnote{Giant molecular
  clouds that pass sufficiently close to a cluster also disrupt it, a
  process first described in the classic work on tidal shocks by
  \citet{spitzer58}.  Passage of the cluster through a spiral arm has
  a similar, impulsive effect \citep{2006A&A...455L..17L}. The latter
  authors find that the two effects together are more efficient than
  the Galactic tidal stripping included in our simulations, at least
  near the solar Galactocentric radius. Hence, our disruption times
  should be considered upper bounds, to be refined by future work.} If
the initial virial radius of a cluster exceeds the tidal radius $r_t$,
as given in equation~(\ref{eq:rt}), then that system disrupts in a few
crossing times, long before either central binary formation or
two-body relaxation can occur. We plot the relation $r_v = r_t$ as the
diagonal, dashed line in Figure~\ref{fig:init}.

This {\em tidal disruption line} represents an extreme limit, as clusters
begin to lose members through the Galactic field well before this
point is reached. In Figure~\ref{fig:tidal}, we show the evolution of
five representative clusters that span the collapse line.  The virial radius $r_v$ plotted is the same as in
Figure~\ref{fig:init}, except that it now represents not the initial
value (here 1.3 pc in all cases), but the instantaneous one that evolves with time.

All curves in Figure~\ref{fig:tidal} initially rise upward,
signifying expansion at constant $N$. Well before reaching the nominal
tidal disruption line, each cluster's members start to be stripped away, and the curve moves to the left.  
Thereafter, each cluster follows a path
roughly parallel to the tidal line but displaced below it. Thus, the
virial radius shrinks with decreasing $N$, but remains a constant
fraction (about 0.5) of the current $r_t$.

We also mark, with a horizontal bar on each evolutionary curve, the
onset of sustained binary formation. From this time forward, there are
one or more hard binaries in the system for most time steps of the
simulation.\footnote{For our purposes, a hard binary is one whose
  internal binding energy exceeds 5 percent of the top-level binding
  energy for the entire cluster.}  In clusters that begin below the
collapse line, the first binaries form very quickly because of mass segregation. The event
is delayed in systems above the collapse line that eventually undergo
core collapse. This delay is caused by the loss of the most massive
cluster members through stellar evolution.

\begin{table}
\begin{centering}
\caption{Characteristic Times of Five Clusters. 
 \label{tab:time}}

\begin{tabular}{lrrrrr}\cline{1-6}
$N$ & $M_{\rm cl}$ & $t_{\rm rel}$ & $t_{\rm bin}$ &  $t_{\rm cc}$ &  $t_{\rm h}$\\
  & ($\msun$) &(Myr) & (Myr) & (Myr)& (Gyr) \\
\cline{1-6}
8,192 & 4,800 &57 & 17 &--- &0.80 \\
16,384 & 9,600 &76 & 24 & ---&1.23 \\
32,768 & 19,000 &112 & 198 & ---&1.82 \\
65,536 & 38,000 &142 & 669 &923 &2.92 \\
131,072 & 77,000 &193 & 1103 & 2131 & {$\cdots$} \\
\cline{1-6}
\end{tabular}
\vspace{.1in}\\
Key evolutionary times for the five clusters shown in
Figure~\ref{fig:tidal}, with $r_v = 1.3\,$pc. The first three columns list the initial
population ($N$), mass ($M_{\rm cl}$), and relaxation time ($t_{\rm rel}$).  The
remaining columns show the times for onset of binary formation
( $t_{\rm bin}$), core collapse ( $t_{\rm cc}$), and tidal stripping of half the initial
mass ( $t_{\rm h}$).
\end{centering}
\end{table}

\begin{figure}
\centering \includegraphics[width=\columnwidth]{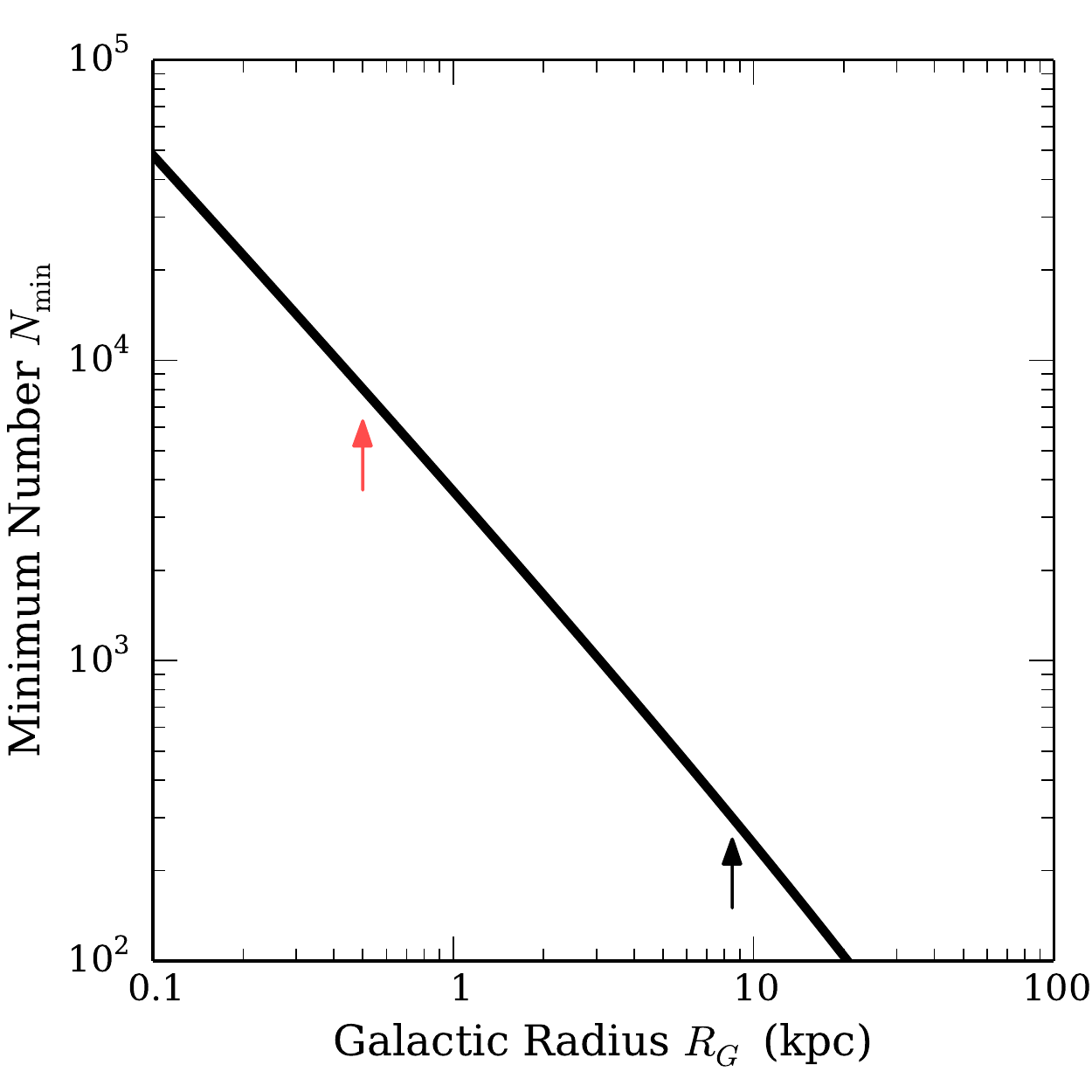}
\caption{  A plot of $N_{\rm min}$, the minimum cluster population
capable of undergoing core collapse, shown as a function of the
Galactocentric radius $R_G$. Most of our simulations were run with $R_G =
8.5\,$kpc,  indicated by the
right vertical arrow. To demonstrate the role of $N_{\rm min}$ explicitly,
we ran two additional simulations with $R_G = 0.5\,$kpc, indicated by the
left vertical arrow.
 \label{fig:nmin}}
\end{figure}

\begin{figure*}
\centering \includegraphics[width=\columnwidth]{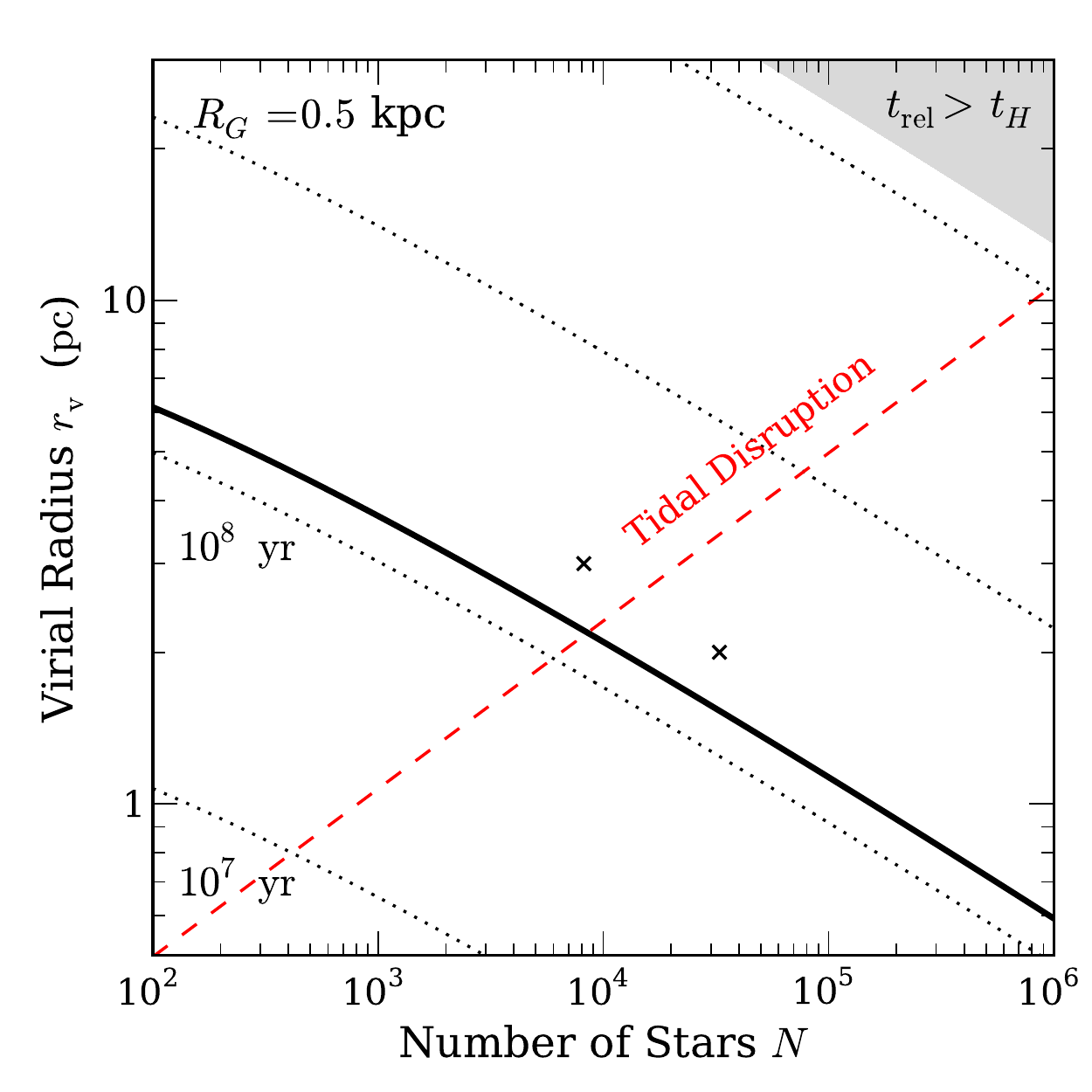} \includegraphics[width=\columnwidth]{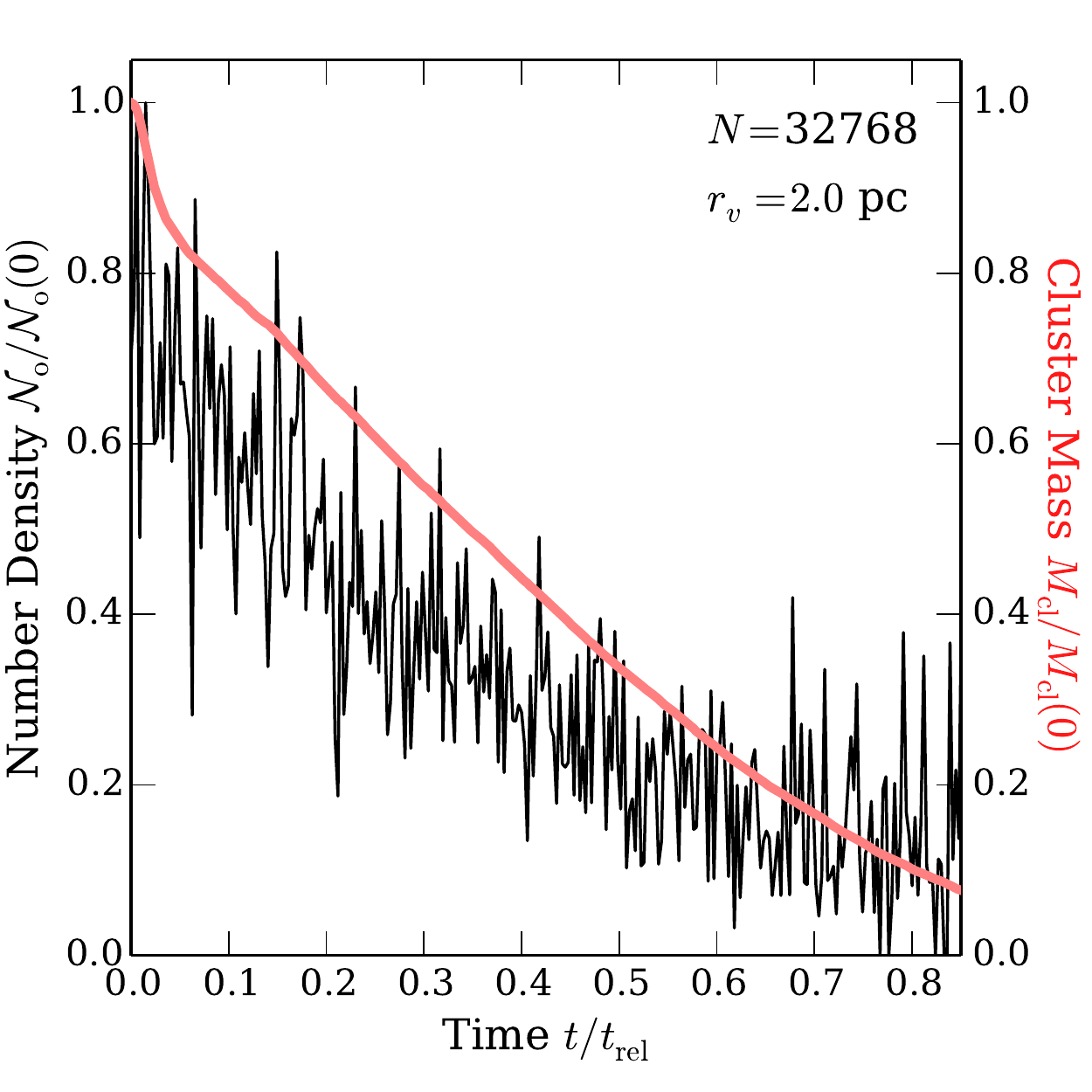}
\caption{ In the left panel, we show the tidal disruption line in the
  $r_v - N$ plane, for $R_G = 0.5$ kpc. We also indicate the initial
  conditions and results of our two simulations using this smaller
  Galactocentric radius. The cluster with $N =$ 8,096 and $r_v = 3.0$
  pc (left cross) disrupts in a few dynamical times. In the right
  panel, we show (thin black line) the evolution of the central
  surface density $\Nc$ for a second cluster that is above the
  collapse line (right cross in the left panel). Its initial
  conditions place this system close enough to the tidal line that it
  undergoes global expansion as it disrupts. We also show (thick red
  line) the decline of the cluster mass $M_{\rm cl}$ due to tidal
  stripping. We again measure time relative to $t_{\rm rel}$, the
  initial relaxation time. \label{fig:nmin2}}
\end{figure*}

In Table~\ref{tab:time}, we list the characteristic evolutionary times for all five
clusters shown in Figure~\ref{fig:tidal}. The first three columns give, respectively, the initial population  $N$,
the total cluster mass, $M_{\rm cl}$, and the relaxation time,  $t_{\rm rel}$, where
the latter was obtained from equation~(\ref{eq:relax}). The time $t_{\rm bin}$ in the fourth
column marks the onset of sustained binary formation at the cluster center.
For the two clusters starting above the collapse line, we next list
$t_{\rm cc}$, the time of core collapse, as judged by the rise in central
density (recall eq.~\ref{eq:ccdef}).  Finally, the last column in the table
gives $t_{\rm h}$, the time by which the Galactic tide has stripped away half
the original mass.\footnote{For the cluster of largest $N$, we stopped the evolution when only 63
percent of the original mass was lost.}

Notice from Table~\ref{tab:time} that central binary formation begins
before core collapse, if the latter occurs at all. Binaries form in
response to the increase in central density, and their heating of the
cluster may or may not prevent a further increase in central
density. Consider, for example, the cluster with initial $r_v =
1.3\,$pc and $N=$32,768, which is below the collapse line. As seen in
Figure~\ref{fig:3x3}, the central density starts to rise at $t \approx
6\,t_{\rm rel}$. Binary heating soon causes the density to fall
again. The cluster with the same $r_v$, but $N=$131,072, starts above
the collapse line. Here, binaries start to form at $t \approx 7 t_{\rm
  rel}$, where Figure~\ref{fig:3x3} shows a relatively small and
transient rise in central density. Later in the evolution, binaries
continue to form, but their heating is insufficient to halt a second,
steeper rise in central density and ultimate core collapse.

Figure~\ref{fig:tidal} shows, as did Figure~\ref{fig:init}, that the tidal disruption and
collapse lines intersect. The cluster population at this intersection,
which we denote $N_{\rm min}$, represents the smallest value for which core
collapse is possible. That is, clusters starting with $N < N_{\rm min}$ simply
expand and then tidally disperse, regardless of their starting virial
radii.

To find a quantitative expression for $N_{\rm min}$, we set ${M_{cl} = N_{\rm min}\mean{m}}$
in equation~(\ref{eq:rt}), and then equate
$r_t$ to the collapse $r_v$ in equation~(\ref{eq:collapse}). We find
\begin{equation}
\label{eq:nmin}
\frac{N_{\rm min}}{\ln( 0.1 N_{\rm min})} =     8.3\,t_{\rm ms}  \frac{m_*}{ \mean{m}} \sqrt{\frac{G M_{\rm enc}}{R_G^3}}.
\end{equation}
It is evident that $N_{\rm min}$ depends on the
Galactocentric radius $R_G$. We show this dependence in Figure~\ref{fig:nmin}.
Although equation~(\ref{eq:nmin}) has no simple analytic solution for $N_{\rm min}$, it is
well fit by a power law:
\begin{equation}
\label{eq:nmineval}
N_{\rm min} \approx 3500 \left(\frac{R_G}{1\,{\rm kpc}}\right)^{-9/8}.
\end{equation} 
Here, we have established the coefficient by using our standard values
for $m_*$, $\mean{m}$, and $t_{\rm ms}$, and by assuming an isothermal potential when
evaluating $M_{\rm enc}$.

If we set $R_G$ equal to the Sun's Galactocentric radius of 8.5 kpc in
equation (\ref{eq:nmineval}), we obtain $N_{\rm min} = 300$, the value seen in Figure~\ref{fig:tidal}, and
indicated by the right vertical arrow in Figure~\ref{fig:nmin}. In order to demonstrate
explicitly the significance of $N_{\rm min}$ through simulations, it is
infeasible to employ this Solar system value of $R_G$, since clusters with $N_{\rm min}
\lesssim 300$ are subject to such large-scale fluctuations that their
central densities do not evolve smoothly. Accordingly, we have rerun
several simulations with $R_G$ lowered to 0.5 kpc, where we expect $N_{\rm min}
\approx 8000$ (see the left vertical arrow in Fig.~\ref{fig:nmin}).

The left panel of Figure~\ref{fig:nmin2} shows the new, shifted tidal disruption
line in the $r_v - N$ plane. Here, $r_v$ has its original meaning from
Figure~\ref{fig:init}, i.e., it is the {\it initial} virial radius. Also indicated
in this panel are the initial conditions for our two simulations, both
with $N$ = 32,768. These conditions place both clusters above the
collapse line, and yet neither actually undergoes core collapse, as the crosses
in the figure indicate.

The cluster starting with $r_v = 3.0$ pc (left cross in
Fig.~\ref{fig:nmin2}) disrupts within a few dynamical times. When
$R_G$ was 8.5 kpc, this same system experienced strong core collapse,
as seen in the top, middle panel of Figure~\ref{fig:3x3}. The cluster
with $r_v = 2.0$ pc also underwent core collapse when $R_G$ was 8.5
kpc (middle panel of Fig.~\ref{fig:3x3}). With the new $R_G$-value, $N
\approx 4 \,N_{\rm min}$, but the system's proximity to the tidal line
leads to a very different evolution. As seen in the right panel of
Figure~\ref{fig:nmin2}, the central surface density monotonically
decreases in less than a relaxation time, as it did when the same cluster was at $R_G = 8.5\,$kpc (see middle panel of Fig.~\ref{fig:3x3}). Now, however,  the total cluster mass $M_{\rm cl}$ falls precipitously in that same interval.
  The
combination of tidal stress and internal mass loss promotes global
expansion until the system completely disperses. By the last time
shown, $t = 0.7\,t_{\rm rel}$, the cluster has lost 90 percent of its
original population.

\section{Discussion}
\label{sec:disc}
\subsection{Summary of results}

In this paper, we have shown that a key process in the theoretical
account of cluster evolution, internal relaxation leading to core
collapse, does not occur for all possible initial conditions.
 Within the $r_v-N$
plane, we have found, first analytically, the collapse line that
separates the two distinct evolutionary paths. For this purpose, we
compared two time scales. The first is the time required for the most
massive stars to settle to the cluster center via dynamical friction.
The second is the main-sequence lifetime of these same stars. Clusters
for which these two times are equal sit on the collapse line in the
$r_v-N$ plane.

Only clusters whose initial sizes and
populations place them above the collapse line evolve in the manner
envisioned by the classical theory, transferring energy outward and
eventually undergoing core collapse with a rapid rise in central
number density. Those starting below the collapse line globally expand
as a result of binary heating that begins before stellar mass loss
drives the cluster to expand.  The central number density of a cluster
born below the collapse line never exceeds its initial value. We have
verified, through a suite of numerical simulations, that clusters
indeed follow these two paths. 
In \S~\ref{sec:predict} and \ref{sec:clusters} below, we use this theoretical
framework, in a preliminary way, to interpret observations of Milky
Way clusters.

For the representative sample of clusters in our study, all eventually disrupt tidally, regardless of where they begin in the $r_v-N$ plane.
 Again, we first proceeded analytically, finding a tidal disruption
line in the  plane. We also tracked the disruption in our simulations, using only the
Galactic potential for simplicity. Finally, we have shown that
clusters below a certain minimum population reach the
 point of tidal disruption without ever undergoing core collapse,
 regardless of their initial size.  Near the Sun's Galactocentric radius, we find that this minimum cluster population is $N_{\rm min} \gtrsim 300$.  Future, more detailed simulations that include tidal
disruption by spiral arms and giant molecular clouds may increase this
figure, although the shape of the tidal line will be similar.

Many of the individual points we have made regarding
cluster dynamics have been described previously. It has long been appreciated that
introducing a stellar mass spectrum dramatically alters the course of
evolution from that of an idealized, uniform-mass system \citep{1974A&A....35..237A}. Similarly,
the critical role of central binaries in both frustrating core
collapse and inducing global expansion is well established \citep{1978RvMP...50..437L}.  That binary heating itself is
inoperative for a cluster that is too large and massive is also known
\citep{1984MNRAS.206..149I,heggiehut,2011MNRAS.410.2787C}, and this fact plays a key role in our evolutionary picture.
 Furthermore, we have stressed the importance of 
disentangling the effects of mass segregation from the phenomenon of
core collapse.

\subsection{Predicting cluster evolution}
\label{sec:predict}

Our theoretical considerations should be useful for gauging the
evolutionary status of real clusters. Drawing the connection is not
entirely straightforward, since we do not observe directly any
cluster's initial state or its evolution through time. The salient questions are the following:
Given a cluster's present-day $r_v$ and $N$, can we determine which
evolutionary path it is on? If the system has not recently undergone
core collapse, which should be apparent observationally, will it do so
in the future? Or will it evolve instead via global expansion?

Answering these questions is easy for clusters presently located {\it
below} the collapse line. All such systems will globally expand until
they begin to be tidally disrupted.  But what about clusters that are currently
{\it above} the line? For these, we first note that there is another
observable, global property of a cluster, its age. This may be
determined, in principle, from the distribution of member stars in the
HR diagram. It is useful to compare this observed age, $t_{\rm obs}$, with a
theoretical ``relative age,'' $t_1$. The latter is the time required for
the cluster to reach its present-day $r_v$ and $N$ starting from the
collapse line. Clusters for which $t_{\rm obs} > t_1$ must have started
below the collapse line, and thus will globally expand in the future.
Here, we are assuming that the cluster is not currently being
disrupted. If it is, then its future history is clear but its prior
history is uncertain, as we shall discuss.

Our simulations allow us to obtain $t_1$ numerically. For a cluster that
starts close to the collapse line, $r_v$ does not change appreciably
until $t \gtrsim 30\,$Myr, at which point stellar mass loss begins to drive
expansion.\footnote{Notice that this period exceeds $t_{\rm ms} = 20\,$Myr,
the main-sequence lifetime of the most massive star. A sizable
fraction of the cluster's mass must be lost for the cluster to begin
expanding.} Over the range of $N$ we have explored for clusters near the collapse line, $r_v$ then increases
as a power law, $r_v \propto t^{0.3}$. Using this relation we find
\begin{equation}
\label{eq:t1}
t_1 \approx 30\,\left(\frac{r_v}{r_{\rm coll}}\right)^{3.3}{\rm Myr},
\end{equation}
where $r_{\rm coll}$ is the virial radius that defines the collapse
line in equation~(\ref{eq:criteval}). 

We previously described the global evolution of clusters starting
above and below the collapse line. In both cases, there is a similar
period of stasis followed by power-law growth (recall Fig.~\ref{fig:expand}).
Eventually, however, the cluster radius peaks and then declines as a
result of tidal stripping. Clusters starting from different initial
conditions can thus traverse the same point in the $r_v-N$ plane. For a
cluster that is actively being disrupted, the relative time $t_1$ is not
useful, and the system's prior history is hard to reconstruct. In some
cases, disruption is evident observationally by the presence of tidal
streamers \citep[e.g.,][]{2003AJ....126.2385O}. The past history of such systems might be
elucidated by studying their internal structure, including the degree
of mass segregation.

\begin{table}
\caption{Open Cluster Sample\label{tab:clusters} }
\begin{tabular}{lrrrr}\cline{1-5}
Name & $N$ & $\sigma$ &  $r_v$ &  Ref \\
\  &  & (km$\,$s$^{-1}$) &  (pc)& \\
\cline{1-5}
Hyades & 550 &  0.30 & 2.5& 1 \\ 
Pleiades (M45) & 800 &  0.36 & 4.4& 2 \\
Praesepe & 800 & 0.67 & 0.87 & 3,4\\ 
NGC 2168 (M35) & 3059 & 0.65 & 2.6 & 5,6 \\ 
NGC 188 & 1050 &0.41 & 10. &7,8 \\

\cline{1-5}

\end{tabular}
\\\vspace{.03in}\\
 {\tiny Refs: 1. \citet{2001A&A...367..111D}, 2. \citet{1998A&A...329..101R}, 3. \citet{2013MNRAS.434.3236K}, 4.\citet{2002A&A...381..446M}, 5. \citet{2012A&A...543A.156K}, 6.\citet{2010AJ....139.1383G}, 7. \citet{2003AJ....126.2922P}, 8. \citet{2008AJ....135.2264G}}
\end{table}

\subsection{Open and globular clusters}
\label{sec:clusters}
Let us now apply these considerations to real systems, starting with
open clusters. At present, the number of open clusters with secure
values of $r_v$ is quite small. The difficulty here is an accurate
determination of the mean velocity dispersion $\sigma$, which is easily
contaminated by binaries for clusters with intrinsically low dispersions. Table~\ref{tab:clusters} shows the modest result of our own
literature search.

In  Figure~\ref{fig:dist}, we plot this handful of open clusters in the
$r_v-N$ plane ({\it squares}), as well as a much larger sample of
globular clusters ({\it filled circles}), to be discussed presently.
Here we have not displayed the tidal disruption line. At least one of
the open clusters (NGC 188) has a Galactocentric radius quite
different from ours, as do most of the globular clusters. In addition,
many of the latter lie well outside the Galactic plane, so that our
approximate form of the gravitational potential (and thus $r_t$) is
inappropriate.

Only two open clusters in our current sample lie above the collapse
line: the Pleiades and NGC 188.
 The former is barely over the line,
and application of equation~(\ref{eq:t1}) yields $t_1 =
40\,$Myr. Since this figure is less than the current age, $t_{\rm obs} =
120\,$Myr \citep{1996ApJ...458..600B}, we conclude that the Pleiades
started below the line and will never undergo core
collapse. \citet{2010MNRAS.405..666C} numerically reconstructed the
history of this system in detail. They indeed found that it began with
a smaller size, $r_v \approx 3\,$pc, and is fated to globally expand until the
point of tidal disruption.

An open cluster much farther above the collapse line is NGC 188. In
this case, we find $t_1 = 800\,$Myr. This number is to be compared with
the empirical age of $t_{\rm obs} = 6.2\,$Gyr \citep{2009AJ....137.5086M}, which makes this one of
the oldest open clusters in the Galaxy. Since $t_{\rm obs} > t_1$, it might
appear once more that the cluster began below the collapse line.
 NGC 188 could have started with $r_v = 2\,$pc and swelled
to its current size in the time $t_{\rm obs}$, assuming its population
remained constant. 
On the other hand,  \citet{2010AJ....139.1889C}  detected an associated tidal streamer. If NGC 188 began
with a significantly higher $N$ and is now disrupting, then its original
location in the $r_v-N$ plane is uncertain, as is its fate.

Let us turn finally to globular clusters. The determination of $r_v$
is now more straightforward, since the relatively small fraction of
binaries cannot appreciably alter the $\sigma$-value observed in the clusters'
spatially unresolved inner regions. In Figure~\ref{fig:dist}, we have placed in
 the $r_v-N$ plane 55 systems whose parameters we obtained from the
\citet[Revision 2010]{1996AJ....112.1487H}~catalog, after assuming a number-to-light ratio of 2.
Virtually all the clusters now lie above the collapse line. However,
it requires further examination to determine their future evolution.

Suppose we set $t_1$ equal to the Hubble time, $t_H$. The
corresponding line lies parallel to and above the collapse line in the
plane. All clusters lying above this line necessarily have $t_{\rm
  obs} < t_1$, and thus will go through core collapse, if they have
not done so already.\footnote{A globular cluster loses members each time it crosses the Galactic
plane. However, these tidal shocks actually accelerate core collapse \citep{1999ApJ...522..935G}.}  One example is 47
Tuc, an especially large and massive cluster with $r_v / r_{\rm coll}
= 8.9$ and $t_1 = 41\,$Gyr. The latter figure naturally exceeds
$t_{\rm obs} = 13\,$Gyr \citep{2010ApJ...708..698D}. A second example,
with nearly the same $t_{\rm obs}$-value, is NGC 7078
(M15). \cite{2000AJ....119.1268G} carefully corrected for the
cluster's rotation to obtain $\sigma$, and we use their figure to find
$t_1 = 26\,$Gyr. In this case, the system has already undergone core
collapse relatively recently \citep{1997ApJ...481..267D}.

Some of the smaller globular clusters that lie closer to the collapse
line may have begun below it. Individual systems require study on a
case by case basis. There is also a clear need to improve the data on
open clusters, so that their evolution can be more fully understood.
So too should the origin and fate of massive systems in the Galactic
plane, such as Westerlund I \citep[for a review, see][]{2010ARA&A..48..431P}.
We leave these tasks to future investigators.

 \begin{figure}
\centering \includegraphics[width=\columnwidth]{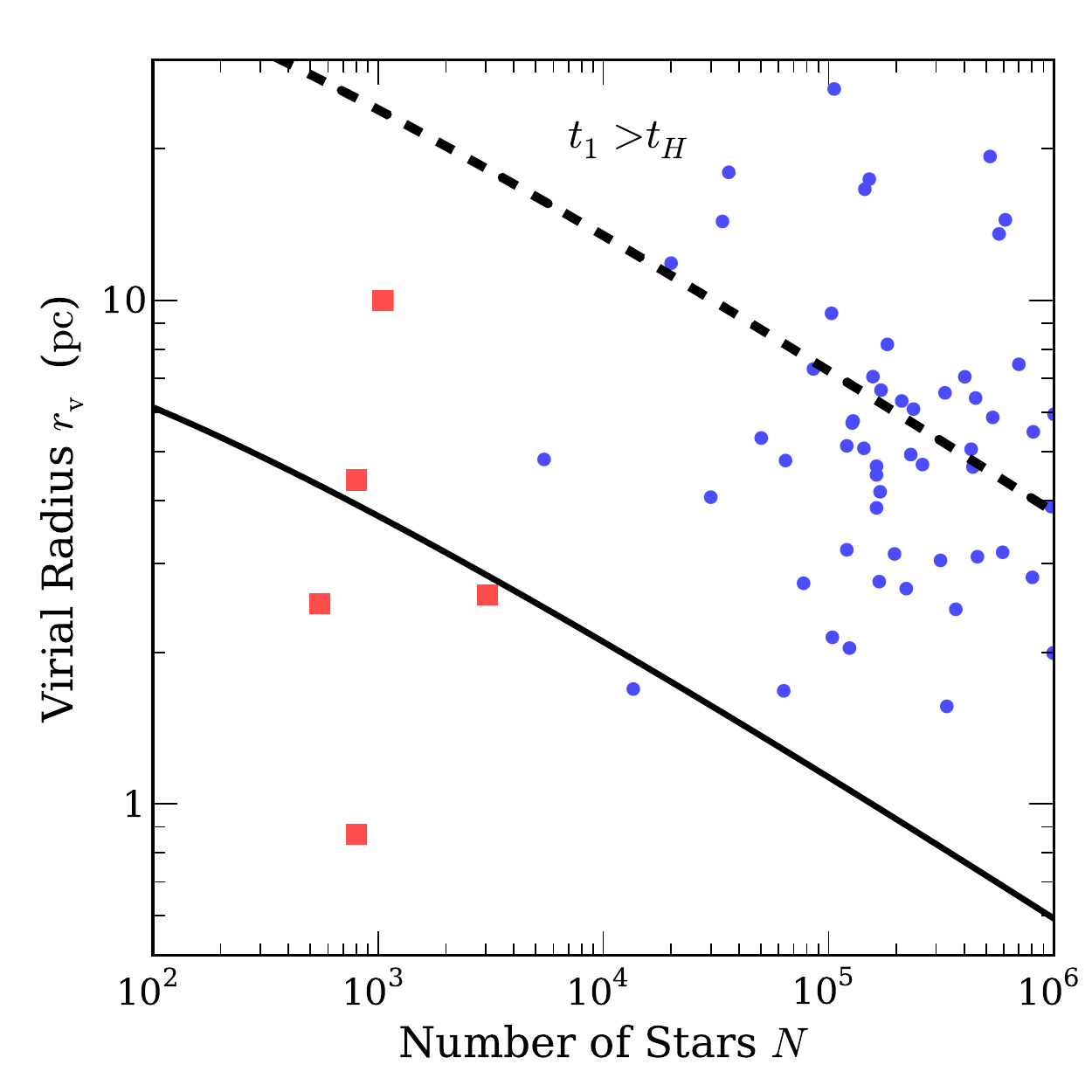}
 \caption{\label{fig:dist} Distribution of observed open clusters ({\it squares}) and globular clusters ({\it circles}) in the $r_v-N$ plane.  Data for the open clusters are collected in Table~\ref{tab:clusters}. Data for the globular clusters were derived from  \citet[Revision 2010]{1996AJ....112.1487H}. The lower, solid curve is the
collapse line. Clusters lying above the upper, dashed curve have $t_1$
exceeding the Hubble time $t_H$. }
 \end{figure}

\section*{Acknowledgments}

 RMO was partially supported by  the Theoretical Astrophysics Center and by the National Aeronautics and Space
 Administration through Einstein Postdoctoral Fellowship Award Number
 PF0-110078 issued by the Chandra X-ray Observatory Center, which is
 operated by the Smithsonian Astrophysical Observatory for and on
 behalf of the National Aeronautics Space Administration under
 contract NAS8-03060.   SS was partially supported by NSF grant 0908573. CPM was partially supported by the Simons Foundation (No. 224959) and NASA NNX11AI97G.

\bibliography{main}

\begin{thebibliography}{65}
\expandafter\ifx\csname natexlab\endcsname\relax\def\natexlab#1{#1}\fi

\bibitem[{{Aarseth}(1971)}]{1971Ap&SS..13..324A}
{Aarseth} S.~J., 1971, \apss, 13, 324

\bibitem[{{Aarseth}(1974)}]{1974A&A....35..237A}
---, 1974, \aap, 35, 237

\bibitem[{{Aarseth}(1999)}]{nbody6}
---, 1999, \pasp, 111, 1333

\bibitem[{{Aarseth}, {H{\'e}non} \& {Wielen}(1974){Aarseth}, {H{\'e}non}, \&
  {Wielen}}]{1974A&A....37..183A}
{Aarseth} S.~J., {H{\'e}non} M., {Wielen} R., 1974, \aap, 37, 183

\bibitem[{{Basri}, {Marcy} \& {Graham}(1996){Basri}, {Marcy}, \&
  {Graham}}]{1996ApJ...458..600B}
{Basri} G., {Marcy} G.~W., {Graham} J.~R., 1996, \apj, 458, 600

\bibitem[{{Baumgardt} \& {Makino}(2003)}]{2003MNRAS.340..227B}
{Baumgardt} H., {Makino} J., 2003, \mnras, 340, 227

\bibitem[{{Bettwieser} \& {Sugimoto}(1984)}]{1984MNRAS.208..493B}
{Bettwieser} E., {Sugimoto} D., 1984, \mnras, 208, 493

\bibitem[{{Binney} \& {Tremaine}(2008)}]{2008gady.book.....B}
{Binney} J., {Tremaine} S., 2008, {Galactic Dynamics: Second Edition}.
  Princeton University Press

\bibitem[{{Casertano} \& {Hut}(1985)}]{core}
{Casertano} S., {Hut} P., 1985, \apj, 298, 80

\bibitem[{{Casetti-Dinescu} {et~al}\mbox{.}(2010){Casetti-Dinescu}, {Girard},
  {Platais}, \& {van Altena}}]{2010AJ....139.1889C}
{Casetti-Dinescu} D.~I., {Girard} T.~M., {Platais} I., {van Altena} W.~F.,
  2010, \aj, 139, 1889

\bibitem[{{Cohn}(1980)}]{cohn1980}
{Cohn} H., 1980, \apj, 242, 765

\bibitem[{{Converse} \& {Stahler}(2010)}]{2010MNRAS.405..666C}
{Converse} J.~M., {Stahler} S.~W., 2010, \mnras, 405, 666

\bibitem[{{Converse} \& {Stahler}(2011)}]{2011MNRAS.410.2787C}
---, 2011, \mnras, 410, 2787

\bibitem[{{de Bruijne}, {Hoogerwerf} \& {de Zeeuw}(2001){de Bruijne},
  {Hoogerwerf}, \& {de Zeeuw}}]{2001A&A...367..111D}
{de Bruijne} J.~H.~J., {Hoogerwerf} R., {de Zeeuw} P.~T., 2001, \aap, 367, 111

\bibitem[{{Djorgovski} \& {King}(1986)}]{1986ApJ...305L..61D}
{Djorgovski} S., {King} I.~R., 1986, \apjl, 305, L61

\bibitem[{{Dotter} {et~al}\mbox{.}(2010){Dotter}, {Sarajedini}, {Anderson},
  {Aparicio}, {Bedin}, {Chaboyer}, {Majewski}, {Mar{\'{\i}}n-Franch}, {Milone},
  {Paust}, {Piotto}, {Reid}, {Rosenberg}, \& {Siegel}}]{2010ApJ...708..698D}
{Dotter} A. {et~al.}, 2010, \apj, 708, 698

\bibitem[{{Dull} {et~al}\mbox{.}(1997){Dull}, {Cohn}, {Lugger}, {Murphy},
  {Seitzer}, {Callanan}, {Rutten}, \& {Charles}}]{1997ApJ...481..267D}
{Dull} J.~D., {Cohn} H.~N., {Lugger} P.~M., {Murphy} B.~W., {Seitzer} P.~O.,
  {Callanan} P.~J., {Rutten} R.~G.~M., {Charles} P.~A., 1997, \apj, 481, 267

\bibitem[{{Fregeau} {et~al}\mbox{.}(2002){Fregeau}, {Joshi}, {Portegies Zwart},
  \& {Rasio}}]{2002ApJ...570..171F}
{Fregeau} J.~M., {Joshi} K.~J., {Portegies Zwart} S.~F., {Rasio} F.~A., 2002,
  \apj, 570, 171

\bibitem[{{Gebhardt} {et~al}\mbox{.}(2000){Gebhardt}, {Pryor}, {O'Connell},
  {Williams}, \& {Hesser}}]{2000AJ....119.1268G}
{Gebhardt} K., {Pryor} C., {O'Connell} R.~D., {Williams} T.~B., {Hesser} J.~E.,
  2000, \aj, 119, 1268

\bibitem[{{Geller} {et~al}\mbox{.}(2010){Geller}, {Mathieu}, {Braden},
  {Meibom}, {Platais}, \& {Dolan}}]{2010AJ....139.1383G}
{Geller} A.~M., {Mathieu} R.~D., {Braden} E.~K., {Meibom} S., {Platais} I.,
  {Dolan} C.~J., 2010, \aj, 139, 1383

\bibitem[{{Geller} {et~al}\mbox{.}(2008){Geller}, {Mathieu}, {Harris}, \&
  {McClure}}]{2008AJ....135.2264G}
{Geller} A.~M., {Mathieu} R.~D., {Harris} H.~C., {McClure} R.~D., 2008, \aj,
  135, 2264

\bibitem[{{Gieles} {et~al}\mbox{.}(2010){Gieles}, {Baumgardt}, {Heggie}, \&
  {Lamers}}]{2010MNRAS.408L..16G}
{Gieles} M., {Baumgardt} H., {Heggie} D.~C., {Lamers} H.~J.~G.~L.~M., 2010,
  \mnras, 408, L16

\bibitem[{{Giersz} \& {Heggie}(1997)}]{1997MNRAS.286..709G}
{Giersz} M., {Heggie} D.~C., 1997, \mnras, 286, 709

\bibitem[{{Gnedin}, {Lee} \& {Ostriker}(1999){Gnedin}, {Lee}, \&
  {Ostriker}}]{1999ApJ...522..935G}
{Gnedin} O.~Y., {Lee} H.~M., {Ostriker} J.~P., 1999, \apj, 522, 935

\bibitem[{{Goodman}(1984)}]{1984ApJ...280..298G}
{Goodman} J., 1984, \apj, 280, 298

\bibitem[{{G{\"u}rkan}, {Freitag} \& {Rasio}(2004){G{\"u}rkan}, {Freitag}, \&
  {Rasio}}]{2004ApJ...604..632G}
{G{\"u}rkan} M.~A., {Freitag} M., {Rasio} F.~A., 2004, \apj, 604, 632

\bibitem[{{Harris}(1996)}]{1996AJ....112.1487H}
{Harris} W.~E., 1996, \aj, 112, 1487

\bibitem[{{Heggie} \& {Hut}(2003)}]{heggiehut}
{Heggie} D., {Hut} P., 2003, {The Gravitational Million-Body Problem: A
  Multidisciplinary Approach to Star Cluster Dynamics}

\bibitem[{{H{\'e}non}(1961)}]{henon1961}
{H{\'e}non} M., 1961, Annales d'Astrophysique, 24, 369

\bibitem[{{H{\'e}non}(1965)}]{1965AnAp...28...62H}
---, 1965, Annales d'Astrophysique, 28, 62

\bibitem[{{Hurley}, {Pols} \& {Tout}(2000){Hurley}, {Pols}, \&
  {Tout}}]{2000MNRAS.315..543H}
{Hurley} J.~R., {Pols} O.~R., {Tout} C.~A., 2000, \mnras, 315, 543

\bibitem[{{Hurley}, {Tout} \& {Pols}(2002){Hurley}, {Tout}, \&
  {Pols}}]{2002MNRAS.329..897H}
{Hurley} J.~R., {Tout} C.~A., {Pols} O.~R., 2002, \mnras, 329, 897

\bibitem[{{Inagaki}(1984)}]{1984MNRAS.206..149I}
{Inagaki} S., 1984, \mnras, 206, 149

\bibitem[{{Ivanova} {et~al}\mbox{.}(2005){Ivanova}, {Belczynski}, {Fregeau}, \&
  {Rasio}}]{2005MNRAS.358..572I}
{Ivanova} N., {Belczynski} K., {Fregeau} J.~M., {Rasio} F.~A., 2005, \mnras,
  358, 572

\bibitem[{{Khalaj} \& {Baumgardt}(2013)}]{2013MNRAS.434.3236K}
{Khalaj} P., {Baumgardt} H., 2013, \mnras, 434, 3236

\bibitem[{{Kharchenko} {et~al}\mbox{.}(2012){Kharchenko}, {Piskunov},
  {Schilbach}, {R{\"o}ser}, \& {Scholz}}]{2012A&A...543A.156K}
{Kharchenko} N.~V., {Piskunov} A.~E., {Schilbach} E., {R{\"o}ser} S., {Scholz}
  R.-D., 2012, \aap, 543, A156

\bibitem[{{Kroupa} \& {Weidner}(2003)}]{2003ApJ...598.1076K}
{Kroupa} P., {Weidner} C., 2003, \apj, 598, 1076

\bibitem[{{Lamers} \& {Gieles}(2006)}]{2006A&A...455L..17L}
{Lamers} H.~J.~G.~L.~M., {Gieles} M., 2006, \aap, 455, L17

\bibitem[{{Larson}(1970)}]{1970MNRAS.147..323L}
{Larson} R.~B., 1970, \mnras, 147, 323

\bibitem[{{Lightman} \& {Shapiro}(1978)}]{1978RvMP...50..437L}
{Lightman} A.~P., {Shapiro} S.~L., 1978, Reviews of Modern Physics, 50, 437

\bibitem[{{Lynden-Bell} \& {Wood}(1968)}]{1968MNRAS.138..495L}
{Lynden-Bell} D., {Wood} R., 1968, \mnras, 138, 495

\bibitem[{{Madsen}, {Dravins} \& {Lindegren}(2002){Madsen}, {Dravins}, \&
  {Lindegren}}]{2002A&A...381..446M}
{Madsen} S., {Dravins} D., {Lindegren} L., 2002, \aap, 381, 446

\bibitem[{{Makino}(1996)}]{makino1996}
{Makino} J., 1996, \apj, 471, 796

\bibitem[{{McMillan}, {Hut} \& {Makino}(1990){McMillan}, {Hut}, \&
  {Makino}}]{1990ApJ...362..522M}
{McMillan} S., {Hut} P., {Makino} J., 1990, \apj, 362, 522

\bibitem[{{Meibom} {et~al}\mbox{.}(2009){Meibom}, {Grundahl}, {Clausen},
  {Mathieu}, {Frandsen}, {Pigulski}, {Narwid}, {Steslicki}, \&
  {Lefever}}]{2009AJ....137.5086M}
{Meibom} S. {et~al.}, 2009, \aj, 137, 5086

\bibitem[{{Meylan} \& {Heggie}(1997)}]{1997A&ARv...8....1M}
{Meylan} G., {Heggie} D.~C., 1997, \aapr, 8, 1

\bibitem[{{Miller} \& {Scalo}(1979)}]{scalo}
{Miller} G.~E., {Scalo} J.~M., 1979, \apjs, 41, 513

\bibitem[{{Nitadori} \& {Aarseth}(2012)}]{GPU}
{Nitadori} K., {Aarseth} S.~J., 2012, \mnras, 424, 545

\bibitem[{{Odenkirchen} {et~al}\mbox{.}(2003){Odenkirchen}, {Grebel}, {Dehnen},
  {Rix}, {Yanny}, {Newberg}, {Rockosi}, {Mart{\'{\i}}nez-Delgado}, {Brinkmann},
  \& {Pier}}]{2003AJ....126.2385O}
{Odenkirchen} M. {et~al.}, 2003, \aj, 126, 2385

\bibitem[{{Platais} {et~al}\mbox{.}(2003){Platais}, {Kozhurina-Platais},
  {Mathieu}, {Girard}, \& {van Altena}}]{2003AJ....126.2922P}
{Platais} I., {Kozhurina-Platais} V., {Mathieu} R.~D., {Girard} T.~M., {van
  Altena} W.~F., 2003, \aj, 126, 2922

\bibitem[{{Plummer}(1911)}]{plummer}
{Plummer} H.~C., 1911, \mnras, 71, 460

\bibitem[{{Portegies Zwart}, {McMillan} \& {Gieles}(2010){Portegies Zwart},
  {McMillan}, \& {Gieles}}]{2010ARA&A..48..431P}
{Portegies Zwart} S.~F., {McMillan} S.~L.~W., {Gieles} M., 2010, \araa, 48, 431

\bibitem[{{Portegies Zwart}, {McMillan} \& {Makino}(2007){Portegies Zwart},
  {McMillan}, \& {Makino}}]{2007MNRAS.374...95P}
{Portegies Zwart} S.~F., {McMillan} S.~L.~W., {Makino} J., 2007, \mnras, 374,
  95

\bibitem[{{Quinlan}(1996)}]{1996NewA....1..255Q}
{Quinlan} G.~D., 1996, \na, 1, 255

\bibitem[{{Raboud} \& {Mermilliod}(1998)}]{1998A&A...329..101R}
{Raboud} D., {Mermilliod} J.-C., 1998, \aap, 329, 101

\bibitem[{{R{\"o}ser} {et~al}\mbox{.}(2010){R{\"o}ser}, {Kharchenko},
  {Piskunov}, {Schilbach}, {Scholz}, \& {Zinnecker}}]{2010AN....331..519R}
{R{\"o}ser} S., {Kharchenko} N.~V., {Piskunov} A.~E., {Schilbach} E., {Scholz}
  R.-D., {Zinnecker} H., 2010, Astronomische Nachrichten, 331, 519

\bibitem[{{Sippel} {et~al}\mbox{.}(2012){Sippel}, {Hurley}, {Madrid}, \&
  {Harris}}]{2012MNRAS.427..167S}
{Sippel} A.~C., {Hurley} J.~R., {Madrid} J.~P., {Harris} W.~E., 2012, \mnras,
  427, 167

\bibitem[{{Spitzer}(1958)}]{spitzer58}
{Spitzer}, Jr. L., 1958, \apj, 127, 17

\bibitem[{{Spitzer}(1969)}]{1969ApJ...158L.139S}
---, 1969, \apjl, 158, L139

\bibitem[{{Tanikawa}, {Hut} \& {Makino}(2012){Tanikawa}, {Hut}, \&
  {Makino}}]{2012NewA...17..272T}
{Tanikawa} A., {Hut} P., {Makino} J., 2012, \na, 17, 272

\bibitem[{{Trager}, {King} \& {Djorgovski}(1995){Trager}, {King}, \&
  {Djorgovski}}]{1995AJ....109..218T}
{Trager} S.~C., {King} I.~R., {Djorgovski} S., 1995, \aj, 109, 218

\bibitem[{{Trenti}, {Heggie} \& {Hut}(2007){Trenti}, {Heggie}, \&
  {Hut}}]{2007MNRAS.374..344T}
{Trenti} M., {Heggie} D.~C., {Hut} P., 2007, \mnras, 374, 344

\bibitem[{{Vesperini}, {McMillan} \& {Portegies Zwart}(2009){Vesperini},
  {McMillan}, \& {Portegies Zwart}}]{2009ApJ...698..615V}
{Vesperini} E., {McMillan} S.~L.~W., {Portegies Zwart} S., 2009, \apj, 698, 615

\bibitem[{{von Hoerner}(1960)}]{1960ZA.....50..184V}
{von Hoerner} S., 1960, \zap, 50, 184

\bibitem[{{Zonoozi} {et~al}\mbox{.}(2011){Zonoozi}, {K{\"u}pper}, {Baumgardt},
  {Haghi}, {Kroupa}, \& {Hilker}}]{2011MNRAS.411.1989Z}
{Zonoozi} A.~H., {K{\"u}pper} A.~H.~W., {Baumgardt} H., {Haghi} H., {Kroupa}
  P., {Hilker} M., 2011, \mnras, 411, 1989

\end{thebibliography}

\end{document}